\documentclass[11pt,draftcls,onecolumn]{IEEEtran}
\IEEEoverridecommandlockouts
\usepackage{amsmath}
\usepackage{color}
\usepackage{graphicx}
\usepackage{latexsym}
\usepackage{epsfig}

\title{On the Controllability and Observability of Networked Dynamic Systems }

\author{Tong Zhou
\thanks{This work was supported in part by the
973 Program under Grant 2009CB320602 and 2012CB316504, the National
Natural Science Foundation of China under Grant 61174122, 61021063,
60721003 and 60625305, and the Specialized Research Fund for the
Doctoral Program of Higher Education, P.R.C., under Grant
20110002110045.}
\thanks{T.Zhou is with the Department of Automation and TNList, Tsinghua University, Beijing, 100084,
CHINA. (Tel: 86-10-62797430; Fax: 86-10-62786911; e-mail:
tzhou@mail.tsinghua.edu.cn.)} }

\begin{document}
\renewcommand{\thefootnote}{\fnsymbol{footnote}}
\maketitle
\renewcommand{\thepage}{30--\arabic{page}}
\setcounter{page}{1}

\begin{abstract}
Some necessary and sufficient conditions are obtained for the controllability and observability of a networked system with linear time invariant (LTI) dynamics. The topology of this system is fixed but arbitrary, and every subsystem is permitted to have different dynamic input-output relations. These conditions essentially depend only on transmission zeros of every subsystem and the connection matrix among subsystems, which makes them attractive in the analysis and synthesis of a large scale networked system. As an application, these conditions are utilized to characterize systems whose steady state estimation accuracy with the distributed predictor developed in \cite{zhou13} is equal to that of the lumped Kalman filter. Some necessary and sufficient conditions on system matrices are derived for this equivalence. It has been made clear that to guarantee this equivalence, the steady state update gain matrix of the Kalman filter must be block diagonal.

\end{abstract}

\IEEEpeerreviewmaketitle

\section{Introduction}

In system analysis and synthesis, a fundamental issue is about verification of its controllability and observability. While the former is concerned with possibilities of maneuvering the internal variables of a system from one state to another state, the latter is concerned with potentials of estimating system internal variables from external measurements. It is now widely known that various important system properties, such as the existence of an optimal control under a criterion like $H_{2}$ or $H_{\infty}$ norm, possibilities of stabilizing a plant and/or locating its poles to a desirable area, convergence of a state estimation procedure like the extensively utilized Kalman filter, etc., are closely related to the controllability and/or observability of the plant at hand \cite{fpp12,ksh,sbkkmpr11,simon06,zdg}. While this issue has been extensively studied over more than half a century and various criteria, such as the rank condition, the PBH test, etc., have been well developed and widely utilized, difficulties arise when these criteria are straightforwardly applied to a system with a large amount of states, which is often encountered in dealing with a networked system consisting of a huge number of subsystems. To be more specific, it has been recognized that when a large scale networked system is to be dealt with, these criteria are usually computationally prohibitive, and various efforts have recently been put in developing a more computationally efficient criterion for this verification \cite{hot11,lsb11,np13,siljak78}.

Owing to extensive pursuits of numerous researchers, significant developments have been achieved in establishing computationally attractive conditions for analyzing  controllability and/or observability of large scale networked systems. For example, in \cite{np13}, Laplacian of a grid graph is adopted in verifying controllability and observability of a family of linear dynamic systems. Through some smart graph decompositions, computable necessary and sufficient conditions are derived which can distinguish all the necessary nodes that lead to a controllable/observable dynamic system. In \cite{lsb11}, on the basis of structural controllability and the cavity method developed in statistical physics, some computationally efficient analytic tools are developed to identify a driver node set that is able to maneuver all the system internal variables. Some interesting results have been obtained from this investigation, which include the observation that driver nodes intend to avoid network hubs, biological regulatory networks are significantly more difficult to be controlled than a social network in the sense that much more driver nodes are required to achieve its controllability, etc.

These efforts have greatly advanced analysis and synthesis of a large scale networked system. Complete settlement of its controllability and/or observability verification problem, however, still requires further investigations, noting that all the existing methods ask some conditions that may not be very easily satisfied by practical problems. For example, the results of \cite{np13} are derived under the condition that every interaction between two connected subsystems has an equal strength, while the approach adopted in \cite{lsb11} requires precise location knowledge on the zero elements of the plant state transition matrix and the controllability is evaluated with a probabilistic metric.

In this paper, we re-investigate controllability and observability of a linear time invariant (LTI) plant consisting of several subsystems. Interactions among subsystems are arbitrary except that the connection matrix is time independent, and the required knowledge is restricted to a state space model of each subsystem and the connection matrix among all subsystems. On the basis of the PBH test, some necessary and sufficient conditions are obtained respectively for the controllability and the observability of the whole system, which depend essentially only on transmission zeros of its subsystems and its connection matrix among all subsystems. This characteristic makes these conditions attractive in analyzing and synthesizing large scale networked systems. As an application of these conditions, situations are discussed under which the distributed state predictor developed in \cite{zhou13} has the same steady state estimation accuracy as that of the lumped Kalman filter. Some necessary and sufficient conditions on system matrices are obtained for this equivalence. It has been made clear that in order to guarantee this equivalence, the update gain matrix of the Kalman filter at its steady state must be block diagonal.

The outline of this paper is as follows. At first, in Section II,
a state space model like representation is given for a networked dynamic system, together with some preliminary results.
Controllability and observability of a networked dynamic system is investigated in Section III. Some necessary and sufficient conditions are given in Section IV for a networked system to have a steady state estimation accuracy with the distributed predictor of \cite{zhou13} equal to that of the Kalman filter. Finally, some concluding remarks are given in Section V in which some further issues are discussed. An appendix is
included to give proofs of some technical results.

The following notation and symbols are adopted. ${\rm\bf
diag}\!\{X_{i}|_{i=1}^{L}\}$ denotes a block diagonal matrix with
its $i$-th diagonal block being $X_{i}$, while ${\rm\bf
col}\!\{X_{i}|_{i=1}^{L}\}$ the vector/matrix stacked by
$X_{i}|_{i=1}^{L}$ with its $i$-th row block vector/matrix being
$X_{i}$. $\{X_{ij}|_{i=1,j=1}^{i=M,j=N}\}$ represents a matrix with
$M\times N$ blocks and its $i$-th row $j$-th column block matrix
being $X_{ij}$, while $0_{m}$ and $0_{m\times n}$ respectively the $m$ dimensional zero column vector and the $m\times n$ dimensional zero matrix. The superscript $T$ and $H$ are used to denote respectively the transpose and the conjugate transpose
of a matrix/vector, and $X^{T}WX$ or $XWX^{T}$ is sometimes
abbreviated as $(\star)^{T}WX$ or $XW(\star)^{T}$, especially when
the term $X$ has a complicated expression. ${\rm\bf E}\!\{\star\}$
is used to denote the mathematical expectation of a random
variable/matrix. When a time dependent function becomes time independent, its temporal variable is usually omitted. For example, when a matrix valued function $A(t)$ degenerates into a constant matrix, it is simply written as $A$.

\section{System Description and Some Preliminaries}

Consider the following networked system ${\rm\bf\Sigma}$ which is constituted from $N$ LTI dynamic subsystems with the dynamics of its $i$-th
subsystem ${\rm\bf\Sigma}_{i}$ being described by
\begin{equation}
\left[\!\!\begin{array}{c} x(t\!+\!1,i) \\ z(t,i) \\
y(t,i)
\end{array}\!\!\right] \!\!=\!\!\left[\!\!\begin{array}{ccc} A_{\rm\bf TT}(i) & A_{\rm\bf TS}(i) & B_{\rm\bf T}(i)  \\
A_{\rm\bf ST}(i) & A_{\rm\bf SS}(i) & B_{\rm\bf S}(i) \\
C_{\rm\bf T}(i) & C_{\rm\bf S}(i) & D_{\rm\bf d}(i)
\end{array}\!\!\right.\! \left.\!\!\begin{array}{c} 0 \\  0 \\
 D_{\rm\bf w}(i)
\end{array}\!\!\right]\! \left[\!\!\begin{array}{c} x(t,i)
\\ v(t,i) \\ d(t,i) \\ w(t,i)
\end{array}\!\!\right]  \label{eqn:7}
\end{equation}
while interactions among its subsystems is described by
\begin{equation}
v(t)=\Phi z(t) \label{eqn:8}
\end{equation}
Here, $z(t)={\rm\bf col}\!\left\{z(t,i)|_{i=1}^{N}\right\}$ and $v(t)={\rm\bf
col}\!\left\{v(t,i)|_{i=1}^{N}\right\}$. Moreover, $t=0,1,\cdots$, and $i=1,2,\cdots,N$, stand respectively for
the temporal variable and the index number of a subsystem, $x(t,i)$ represents the state vector of the
$i$-th subsystem ${\rm\bf\Sigma}_{i}$ at time $t$, $z(t,i)$ and
$v(t,i)$ respectively its output vector to other subsystems and
input vector from other subsystems, $y(t,i)$, $d(t,i)$ and $w(t,i)$
respectively its output vector, input/process disturbance vector
and measurement error vector. To distinguish the output vector $z(t,i)$ and the input vector $v(t,i)$ respectively from the output vector $y(t,i)$ and the input vector $d(t,i)$, $z(t,i)$ and $v(t,i)$ are called internal output/input vectors, while $y(t,i)$ and $d(t,i)$ external output/input vectors.

In the above system description, every subsystem of the plant is permitted to have different dynamics, but their dynamics is required to be time invariant. On the other hand, the relation of Equation (\ref{eqn:8})
reflects the fact that the internal inputs of a subsystem are actually constituted from and only from some internal outputs of other subsystems. As only time invariant systems are investigated in this paper, connections among plant subsystems are also time invariant, which means that the matrix $\Phi$ is a constant matrix. On the other hand, note that large scale systems are usually sparse \cite{ad03,siljak78,zhou13}. These observations imply that every row of the matrix $\Phi$ has only one nonzero element which is equal to one,  and compared to the number of the states of the whole system, the dimension of this matrix is in general not high \cite{ad03,zhou09,zhou13}.

It is worthwhile to point out that the restriction that all the nonzero elements of the connection matrix $\Phi$ are equal to $1$ does not mean that influence strengthes among all subsystems are equivalent to each other. In fact, different subsystem influence strengthes can be reflected in both the connection matrix $\Phi$ and the subsystem parameter matrices like $A_{\rm\bf TS}(i)$, $A_{\rm\bf SS}(i)$, etc. \cite{ad03,zhou09,zhou13}. In this paper, in order to reduce computational complexity in verifying controllability and/or observability of the dynamic system $\rm\bf\Sigma$, influence strengthes among subsystems are selected to be included in its subsystem parameter matrices.

The above model is a modification of the model adopted in describing the dynamics of a spatially connected plant which was originally suggested in \cite{ad03} and utilized in many other studies such as \cite{zhou09, zhou13}. The differences between these two models are that the model of Equations (\ref{eqn:7}) and (\ref{eqn:8}) permits its subsystems to have different dynamics and connections among subsystems to be arbitrary. This makes the model adopted in this paper capable of describing dynamics of a larger class of physical systems. For example, in addition to a finite element approximation to distributed parameter systems, it is also able to describe power networks, multi-agent systems, etc. The model given by Equations (\ref{eqn:7}) and (\ref{eqn:8}) is very similar to the state space representation of a plant, but knowledge on subsystem connections are described more explicitly. To avoid confusions, it is sometimes called a state space model like representation \cite{zhou09,zhou13}.

The above model can be easily modified to systems having time varying dynamics and time varying connections. Some other well encountered factors, such as communication  delays, data missing, etc., can also be introduced into this system model. However, as controllability and observability verification for time varying systems is in general still a mathematically difficult problem in system theories, and time delay and packet loss, etc., will lead to other awkward issues which are currently under investigations, we adopt the above system model in this paper as the first step towards investigations on networked systems.

To investigate controllability and observability of the system described by Equations (\ref{eqn:7}) and (\ref{eqn:8}), the following results are required, which are widely known as the PBH test and have been extensively applied in control and system theories, and can be found in various textbooks dealing with state estimations and/or controller designs, for example, \cite{ksh,zdg}.

\hspace*{-0.4cm}{\bf Lemma 1.} Assume that a discrete LTI system has the following dynamic input-output relation,
\begin{equation}
x(t+1)=Ax(t)+Bd(t),\hspace{0.5cm}
y(t)=Cx(t)+Dw(t)\label{eqn:9}
\end{equation}
in which $A$, $B$, $C$ and $D$ are constant matrices with compatible dimensions, $x(t)$ is the plant state vector, $d(t)$ and $w(t)$ are the plant external input vectors. Then,
\begin{itemize}
\item The system is controllable if and only if for every complex scalar $\lambda$ and every nonzero complex vector $x$ satisfying $x^{H}A=\lambda x^{H}$, $x^{H}B\neq 0$.
\item The system is observable if and only if for every complex scalar $\lambda$ and every nonzero complex vector $y$ satisfying $Ay=\lambda y$, $Cy\neq 0$.
\end{itemize}

The next lemma provides some characteristics of a plant transmission zero, which is closely related to the existence of a nonzero plant input that makes its output constantly equal to zero. This result is also widely known in linear system theories \cite{zdg}.

\hspace*{-0.4cm}{\bf Lemma 2.} Let $G(\lambda)$ be a proper transfer function matrix (TFM) having full column normal rank. Then, a complex number $\lambda_{0}$ is a transmission zero of this TFM if and only if there exists a nonzero complex vector $z_{0}$ satisfying $G(\lambda_{0})z_{0}=0$.

\section{Controllability and Observability of a Networked System}

As mentioned before, system controllability and/or observability have been extensively investigated for a long time and various approaches have been developed, such as the rank condition, the PBH test, etc.. It is, however, still a challenging issue for a large scale system, noting that when the existing approaches are directly utilized, some related matrices usually have a very high dimension that makes actual verification computationally prohibitive. On the other hand, when the networked system of Equations (\ref{eqn:7}) and (\ref{eqn:8}) is to be analyzed, it is sometimes not a very easy task even to obtain a state space representation of the whole system itself, especially when the dimension of the matrix $\Phi$ is large which corresponds to dense interactions among subsystems. In fact, developing computationally efficient methods for validating controllability and/or observability of a large scale networked system is currently a very active research topic, and many significant advances have been achieved for some specific systems \cite{lsb11,np13}.

To make mathematical derivations more concise, the following
matrices are at first defined. $A_{\rm\bf
*\#}\!\!=\!\!{\rm\bf diag}\!\left\{\!A_{\rm\bf
*\#}(i)|_{i=1}^{N}\!\right\}$, $B_{\rm\bf T}\!\!=\!\!{\rm\bf
diag}\!\!\left\{\!B_{\rm\bf T}(i)|_{i=1}^{N}\!\right\}$, $B_{\rm\bf S}\!\!=\!\!{\rm\bf
diag}\!\!\left\{\!B_{\rm\bf S}(i)|_{i=1}^{N}\!\right\}$,
$C_{\rm\bf T}\!\!=\!\!{\rm\bf diag}\!\!\left\{\!C_{\rm\bf
T}(i)|_{i=1}^{N}\!\right\}$, $C_{\rm\bf S}\!=\!{\rm\bf
diag}\!\left\{C_{\rm\bf S}(i)|_{i=1}^{N}\right\}$, $D_{\rm\bf d}\!=\!{\rm\bf diag}\!\left\{D_{\rm\bf d}(i)|_{i=1}^{N}\right\}$ and
$D_{\rm\bf w}\!=\!{\rm\bf diag}\!\left\{D_{\rm\bf w}(i)|_{i=1}^{N}\right\}$, in which
${\rm\bf *,\#}={\rm\bf T}$ or ${\rm\bf S}$. Moreover, denote
${\rm\bf col}\!\left\{d(t,i)|_{i=1}^{N}\right\}$, ${\rm\bf
col}\!\left\{w(t,i)|_{i=1}^{N}\right\}$, ${\rm\bf
col}\!\left\{x(t,i)|_{i=1}^{N}\right\}$ and ${\rm\bf
col}\!\left\{y(t,i)|_{i=1}^{N}\right\}$ respectively by
$d(t)$, $w(t)$, $x(t)$ and $y(t)$. Then, straightforward
algebraic manipulations show that when the dynamic system ${\rm\bf \Sigma}$ is well-posed, which is equivalent to the regularity of the matrix $I-A_{\rm\bf SS}\Phi$, its  dynamics can be equivalently described by the following state space
representation
\begin{equation}
\hspace*{-0.8cm}\left[\!\!\begin{array}{c} x(t\!+\!1) \\  y(t)
\end{array}\!\!\right] \!\!=\!\!\left\{\!\left[\!\!\begin{array}{ccc}
A_{\rm\bf TT} & \hspace*{-0.2cm} B_{\rm\bf T} & \hspace*{-0.2cm} 0  \\
C_{\rm\bf T}  & \hspace*{-0.2cm} D_{\rm\bf d} & \hspace*{-0.2cm} D_{\rm\bf w}
\end{array}\!\!\right]\!\!+\!\!\left[\!\!\begin{array}{c}
A_{\rm\bf TS} \\
C_{\rm\bf S}\end{array}\!\!\right]\!\!\Phi\!
\left[\;I\!-\!A_{\rm\bf SS}\Phi\;\right]^{\!-1}\!\left[
A_{\rm\bf ST}\;\; B_{\rm\bf S}\;\;
0\right]\right\}\!\!\left[\!\!\begin{array}{c} x(t)
\\ d(t) \\ w(t)
\end{array}\!\!\right]  \label{eqn:10} \\
\end{equation}

Note that well-posedness is an essential property required in system designs. In fact, a plant that is not well-posed is usually hard to control and/or unable to estimate \cite{ksh,simon06,zdg}. It is therefore assumed throughout this paper that all the systems are well-posed. This means that the inverse of the matrix $I-A_{\rm\bf SS}\Phi$ always exists.

Define matrices $A$, $B$, $C$ and $D$ respectively as $A\!=\!A_{\rm\bf
TT}\!+\!A_{\rm\bf TS}\Phi\left[I\!-\!A_{\rm\bf
SS}\Phi\right]^{-1}A_{\rm\bf ST}$, $B\!=\!B_{\rm\bf
T}\!+\!A_{\rm\bf TS}\Phi\left[I\!-\!A_{\rm\bf
SS}\Phi\right]^{-1}B_{\rm\bf S}$,  $C=C_{\rm\bf T}+C_{\rm\bf S}\Phi\left[\;I-A_{\rm\bf
SS}\Phi\;\right]^{-1}\!A_{\rm\bf ST}$ and $D\!=\left[\!D_{\rm\bf
d}\!+\!C_{\rm\bf S}\Phi\left[I\!-\!A_{\rm\bf
SS}\Phi\right]^{\!-1}\!\!B_{\rm\bf S}\;\;D_{\rm\bf w}\right]$. Clearly, all these matrices are time invariant. Moreover, the input-output relation of the LTI dynamic system ${\rm\bf\Sigma}$ can be further simplified to
\begin{displaymath}
x(t+1)=Ax(t)+Bd(t),\hspace{0.5cm} y(t)=Cx(t)+D{\rm\bf col}\{d(t),\;w(t)\}
\end{displaymath}
which is very similar to that of Equation (\ref{eqn:9}), and makes Lemma 1 applicable to the analysis of the controllability and observability of the dynamic system ${\rm\bf\Sigma}$.

In this section, observability of the dynamic system ${\rm\bf \Sigma}$ is at first investigated. Its controllability is discussed afterwards through dualities between controllability and observability. For this purpose, the following symbols are introduced. Assume that the dimensions of
$x(t,i)$ and $v(t,i)$ are respectively $m_{{\rm\bf T}i}$ and
$m_{{\rm\bf S}i}$. Define integers $M_{{\rm\bf T}i}$, $M_{{\rm\bf
S}i}$, $M_{\rm\bf T}$ and $M_{\rm\bf S}$ as $M_{\rm\bf
T}={\sum_{k=1}^{N} m_{{\rm\bf T}k}}$, $ M_{\rm\bf S}={\sum_{k=1}^{N}
m_{{\rm\bf S}k}}$, and $M_{{\rm\bf T}i}=M_{{\rm\bf S}i}=0$ when
$i=1$, $M_{{\rm\bf T}i}={\sum_{k=1}^{i-1} m_{{\rm\bf T}k}}$,
$M_{{\rm\bf S}i}={\sum_{k=1}^{i-1} m_{{\rm\bf S}k}}$ when $2\leq
i\leq N$.

From the aforementioned relations and Lemma 1, the following result is obtained about the observability of the dynamic system ${\rm\bf\Sigma}$, while its proof is deferred to the appendix.

\hspace*{-0.4cm}{\bf Theorem 1.} Assume that the dynamic system ${\rm\bf\Sigma}$ is well-posed. Then, this dynamic system is observable if and only if for every complex scalar $\lambda$, the matrix valued polynomial (MVP) $M(\lambda)$, which is defined in the following equation, is of full column rank.
\begin{equation}
M(\lambda)=\left[\begin{array}{cc}
\lambda I_{M_{\rm\bf T}}-A_{\rm\bf TT} & -A_{\rm\bf TS} \\
-C_{\rm\bf T} & -C_{\rm\bf S} \\
-\Phi A_{\rm\bf ST} & I_{M_{\rm\bf S}}-\Phi A_{\rm\bf SS} \end{array}\right]     \label{eqn:14}
\end{equation}

Theorem 1 makes it clear that observability of the dynamic system ${\rm\bf\Sigma}$ can be investigated without a state space model of the whole system. This is an interesting  observation as sometimes state space models are not very easy to be established for a networked system, especially when its scale is large. However, to make its results implementable, some computationally more attractive conditions must be developed.

The next lemma gives an equivalent condition on the rank deficiency of the aforementioned MVP.

\hspace*{-0.4cm}{\bf Lemma 3.} Define a TFM $G(\lambda)$ as
\begin{equation}
G(\lambda)=\left[\begin{array}{c}
C_{\rm\bf S} \\ \Phi A_{\rm\bf SS}-I_{M_{\rm\bf S}} \end{array}\right]+
\left[\begin{array}{c}
C_{\rm\bf T} \\ \Phi A_{\rm\bf ST} \end{array}\right]
\left(\lambda I_{M_{\rm\bf T}}-A_{\rm\bf TT}\right)^{-1}A_{\rm\bf TS}
\label{eqn:15}
\end{equation}
Then, the MVP $M(\lambda)$ defined in Equation (\ref{eqn:14}) does  not always have a full column rank if and only if the TFM $G(\lambda)$ has a transmission zero.

A proof of this lemma is given in the appendix.

Note that except the connection matrix $\Phi$, all the other matrices in the definition of the TFM $G(\lambda)$ have a block diagonal structure which is consistent with  each other. This particular structure enables developments of computationally efficient procedures for verifying observability of the dynamic system ${\rm\bf\Sigma}$.

Define TFMs $G^{[1]}(\lambda)$, $G^{[2]}(\lambda)$, $G^{[1]}_{i}(\lambda)$ and $G^{[2]}_{i}(\lambda)$, $i=1,2,\cdots,N$, respectively as
\begin{eqnarray*}
& & G^{[1]}(\lambda)={\rm\bf diag}\left\{\left.G^{[1]}_{i}(\lambda)\right|_{i=1}^{N}\right\},\hspace{0.5cm}
G^{[2]}(\lambda)={\rm\bf diag}\left\{\left.G^{[2]}_{i}(\lambda)\right|_{i=1}^{N}\right\} \\
& & G^{[1]}_{i}(\lambda)=C_{\rm\bf S}(i)+C_{\rm\bf T}(i)\left[\lambda I_{m_{\rm\bf Ti}}-A_{\rm\bf TT}(i)\right]^{-1}A_{\rm\bf TS}(i)  \\
& & G^{[2]}_{i}(\lambda)=A_{\rm\bf SS}(i)+A_{\rm\bf ST}(i)\left[\lambda I_{m_{\rm\bf Ti}}-A_{\rm\bf TT}(i)\right]^{-1}A_{\rm\bf TS}(i)
\end{eqnarray*}
Assume that the TFMs $G^{[1]}(\lambda)$ and $G^{[1]}_{i}(\lambda)$ have respectively $m$ and  $m_{i}$ distinctive transmission zeros, $i=1,2,\cdots,N$. Then, from Lemma 2 and
$G^{[1]}(\lambda)={\rm\bf diag}\{G^{[1]}_{i}(\lambda)|_{i=1}^{N}\}$, it can be proved that for each $i=1,2,\cdots,N$, every transmission zero of the TFM $G^{[1]}_{i}(\lambda)$ is also a transmission zero of the TFM $G^{[1]}(\lambda)$. As different TFMs may share some common transmission zeros, it is clear that $\max_{1\leq i\leq N}m_{i}\leq m\leq \sum_{i=1}^{N}m_{i}$.

Let $\lambda_{0}^{[k]}$, $k=1,2,\cdots,m$, be a transmission zero of the TFM $G^{[1]}_{k(s)}(\lambda)$, $s=1,2,\cdots,s^{[k]}$ with $1\leq s^{[k]}\leq N$, $k(1)<k(2)<\cdots<k(s^{[k]})$, and $k(s)\in \{1,\;2,\;\cdots,\;N\}$. Moreover, let ${\cal Y}^{[k]}_{s}$ denote the set consisting of all complex vectors $y_{s,0}^{[k]}$ satisfying $G^{[1]}_{k(s)}(\lambda_{0}^{[k]})y_{s,0}^{[k]}=0$. Using these vector sets, construct another vector set ${\cal Y}^{[k]}$ as
\begin{equation}
\hspace*{-0.5cm} {\cal Y}^{[k]}\!=\!\left\{y\left|\begin{array}{l}
y\!=\!{\rm\bf col}\!\left\{\!\!\left.\left(0_{m_{{\rm\bf S}(k(i)+1)}},\;\cdots,\;0_{m_{{\rm\bf S}(k(i+1)-1)}},\; y_{i+1,0}^{[k]}\right)\right|_{i=0}^{s^{[k]}-1},\;
0_{m_{{\rm\bf S}(k(s^{[k]})+1)}},\;\cdots,\;0_{m_{{\rm\bf S}N}}\!\right\} \\
\hspace*{2cm} y_{i,0}^{[k]}\in{\cal Y}^{[k]}_{i},\;i=1,2,\cdots,s^{[k]};\;\; y\neq 0 \end{array}\!\! \right.\!\!\right\}
\end{equation}
in which $k(0)$ is defined as $k(0)=0$. Let $m_{{\rm\bf y}i}$, $i=1,2,\cdots,N$, stand for the dimension of the external output vector $y(t,i)$ of the plant's $i$-th subsystem ${\rm\bf\Sigma}_{i}$. Then, from the definition of the set ${\cal Y}^{[k]}$, it can be directly proved that for every $y^{[k]}\in{\cal Y}^{[k]}$,
\begin{eqnarray}
G^{[1]}(\lambda_{0}^{[k]})y^{[k]}\!\!\!\!&=&\!\!\!\!{\rm\bf diag}\left\{\left.G^{[1]}_{i}(\lambda^{[k]}_{0})\right|_{i=1}^{N}\right\}y^{[k]} \nonumber\\
& &\hspace*{-1.5cm}=\!{\rm\bf col}\!\left\{\!\!\left.\left(0_{m_{{\rm\bf y}(k(i)+1)}},\;\cdots,\;0_{m_{{\rm\bf y}(k(i+1)-1)}},\; G^{[1]}_{k(i+1)}(\lambda^{[k]}_{0})y_{i+1,0}^{[k]}\right)\right|_{i=0}^{s^{[k]}-1},\;
0_{m_{{\rm\bf y}(k(s^{[k]})+1)}},\;\cdots,\;0_{m_{{\rm\bf y}N}}\!\right\}\nonumber\\
& &\hspace*{-1.5cm}=\! 0
\end{eqnarray}

On the basis of Theorem 1 and Lemmas 2 and 3, a necessary and sufficient condition can be derived for the observability of the dynamic system ${\rm\bf\Sigma}$ that can lead to a computationally efficient verification procedure.

\hspace*{-0.4cm}{\bf Theorem 2.} The dynamic system ${\rm\bf\Sigma}$ is observable if and only if for every $k\in\{1,2,\cdots,m\}$ and every $y^{[k]}\in{\cal Y}^{[k]}$, $\Phi G^{[2]}(\lambda_{0}^{[k]})y^{[k]}\neq y^{[k]}$.

A proof of this theorem can be found in the appendix.

Recall that existence of transmission zeros in a plant TFM means the existence of a nonzero input which can make the plant output constantly equal to zero \cite{zdg}. Note that the TFMs $G_{i}^{[1]}(\lambda)$ and $G_{i}^{[2]}(\lambda)$ reflecting influences of the internal input vector $v(t,i)$ respectively on the external output $y(t,i)$ and the internal output $z(t,i)$ when the subsystem ${\rm\bf\Sigma}_{i}$ is isolated from other subsystems. The conditions of Theorem 2 can therefore be explained that if influences of some internal inputs of the subsystem ${\rm\bf\Sigma}_{i}$ on its state vector are not reflected in its external output vector, then, in order to guarantee possibilities of estimating these influences, there must exist some output signals that can reflect these influences and be transmitted to other subsystems of the plant. In other words, if some of a subsystem's states can not be estimated in some particular situations from its own measurements, these states must have effects on other subsystems which enables their estimations. From this viewpoint, it can be claimed that conditions of Theorem 2 have some clear physically significant  interpretations.

An interesting property of the above conditions on system observability is that while they depends on parameter matrices of every subsystem, they do not depend on the parameter matrices of the whole networked system. This is attractive in dealing with a plant whose subsystem models are much easier to obtain than the whole plant. Such a situation is not very rare in engineering applications \cite{ad03,siljak78,zhou13}.

Concerning actual implementations, note that there are various efficient methods for calculating the transmission zeros of a TFM, as well as their related vectors. For example, using the McMillan form of a TFM that can be obtained through elementary algebraic manipulations, its transmission zeros and the related vectors can be straightforwardly obtained. It is also possible to compute the transmission zeros of a square system using its inverse TFM \cite{zdg}. On the other hand, note that the TFM $G_{i}^{[1]}(\lambda)$ depends only on the state space representation like model of the $i$-th subsystem of the plant. This means that $G_{i}^{[1]}(\lambda)$ is generally not very difficult to compute, as subsystems of a networked system, even if the system itself is of a large scale, usually have a low dimension for all its state vector, input vector and output vector. It is also worthwhile to mention the fact that every row vector of the matrix $\Phi$ only has one nonzero element which is usually equal to $1$ and a large scale system usually has a sparse structure. This means that if the vector $y^{[k]}$ is partitioned as $y^{[k]}={\rm\bf col}\{y_{i}^{[k]}|_{i=1}^{N}\}$ with $y_{i}^{[k]}$ a $m_{{\rm\bf S}i}$ dimensional column vector, the computation $\Phi G^{[2]}(\lambda_{0}^{[k]})y^{[k]}$ is essentially equivalent to that of $G^{[2]}_{k(s)}(\lambda_{0}^{[k]})y_{k(s)}^{[k]}$, $s=1,2,\cdots,s^{[k]}$, which also depends only on the parameter matrices of the plant subsystem ${\rm\bf \Sigma}_{k(s)}$.

Assume that the dimension of the internal output vector $z(t,i)$ is $m_{{\rm\bf z}i}$, $i=1,2,\cdots,N$. Define integers $M_{{\rm\bf y}i}$ and $M_{{\rm\bf
z}i}$, $M_{\rm\bf y}$ and $M_{\rm\bf z}$ respectively as $M_{\rm\bf
y}={\sum_{k=1}^{N} m_{{\rm\bf y}k}}$, $ M_{\rm\bf z}={\sum_{k=1}^{N}
m_{{\rm\bf z}k}}$, and $M_{{\rm\bf y}i}=M_{{\rm\bf z}i}=0$ when
$i=1$, $M_{{\rm\bf y}i}={\sum_{k=1}^{i-1} m_{{\rm\bf y}k}}$,
$M_{{\rm\bf z}i}={\sum_{k=1}^{i-1} m_{{\rm\bf z}k}}$ when $2\leq
i\leq N$. Let $y_{s,i}^{[k]}$, $i=1,2,\cdots,p(k,s)$, denote all the basis vectors of the null space of $G^{[1]}_{k(s)}(\lambda_{0}^{[k]})$, in which $1\leq s\leq s^{[k]}$ and $1\leq k\leq m$. Define a matrix $Y_{s}^{[k]}$ as $Y_{s}^{[k]}=\left[y_{s,1}^{[k]},\; y_{s,2}^{[k]},\;\cdots,\;y_{s,p(k,s)}^{[k]}\right]$. Then, for every nonzero $y_{s,0}^{[k]}$ satisfying $G^{[1]}_{k(s)}(\lambda_{0}^{[k]})y_{s,0}^{[k]}=0$, there exists one and only one nonzero vector $\alpha_{s}^{[k]}$, such that $y_{s,0}^{[k]}=Y_{s}^{[k]}\alpha_{s}^{[k]}$ \cite{hj91}. Define matrix $Z_{s}^{[k]}$ as $Z_{s}^{[k]}=G^{[2]}_{k(s)}(\lambda_{0}^{[k]})Y_{s}^{[k]}$.
Then, it can be straightforwardly proved that if the vector $y^{[k]}$ belongs to the set ${\cal Y}^{[k]}$, then, there exists a $\sum_{s=1}^{s^{[k]}}p(k,s)$ dimensional column vector $\alpha^{[k]}$, such that $y^{[k]}$ is not equal to zero if and only if $\alpha^{[k]}\neq 0$, and
\begin{equation}
y^{[k]}=Y^{[k]}\alpha^{[k]},\hspace{0.5cm} G^{[2]}(\lambda_{0}^{[k]})y^{[k]}=Z^{[k]}\alpha^{[k]}
\end{equation}
in which
\begin{eqnarray}
& & Y^{[k]}=\left[\left.{\rm\bf col}\left\{0_{M_{{\rm\bf y}k(i)}\times p(k,i)},\; Y_{i}^{[k]},\; 0_{(M_{\rm\bf y}-M_{{\rm\bf y}(k(i)+1)})\times p(k,i)}\right\}\right|_{i=1}^{s^{[k]}}\right] \\
& & Z^{[k]}=\left[\left.{\rm\bf col}\left\{0_{M_{{\rm\bf z}k(i)}\times p(k,i)},\; Z_{i}^{[k]},\; 0_{(M_{\rm\bf z}-M_{{\rm\bf z}(k(i)+1)})\times p(k,i)}\right\}\right|_{i=1}^{s^{[k]}}\right]
\end{eqnarray}


Therefore, $\Phi G^{[2]}(\lambda_{0}^{[k]})y^{[k]}-y^{[k]}=(\Phi Z^{[k]}-Y^{[k]})\alpha^{[k]}$. Hence, the nonexistence of a $y^{[k]}$ simultaneously satisfying $y^{[k]}\in{\cal Y}^{[k]}$ and $\Phi G^{[2]}(\lambda_{0}^{[k]})y^{[k]}=y^{[k]}$ is equivalent to that the matrix $\Phi Z^{[k]}-Y^{[k]}$ is of full column rank. Clearly from their definitions, the matrices $\Phi$, $Y^{[k]}$ and $Z^{[k]}$ usually have a sparse structure. Note that sparse matrices are a well investigated matrix set and there are various efficient methods dealing with computations and/or verifying properties of this type of matrices, such as those reported in \cite{ggl93}. It can be claimed that computationally efficient algorithm can be developed for verifying whether or not the matrix $\Phi Z^{[k]}-Y^{[k]}$ is of full column rank.

Based on these results, a prototypical procedure can be constructed for verifying the observability of the dynamic system $\rm\bf\Sigma$.

{\rm\bf Observability Verification Algorithm:}
\begin{enumerate}
\item Compute all transmission zeros of the TFM $G^{[1]}_{i}(\lambda)$, $i=1,2,\cdots,N$. Construct the set $\{\lambda_{0}^{[1]},\;\lambda_{0}^{[2]},\;$ $\cdots,\;\lambda_{0}^{[m]}\}$ which consists of all their distinctive values.
    Assign the index $k$ as $k=1$.
\item For a fixed $k$ satisfying $1\leq k\leq m$, construct the matrix $Y_{s}^{[k]}$ constituting from all the basis vectors of the null space of
    $G^{[1]}_{k(s)}(\lambda_{0}^{[k]})$ for each $1\leq s\leq s^{[k]}$.
\item Compute $Z_{s}^{[k]}$ as $Z_{s}^{[k]}=G^{[2]}_{k(s)}(\lambda_{0}^{[k]})Y_{s}^{[k]}$, $1\leq s\leq s^{[k]}$. Construct matrices $Y^{[k]}$ and $Z^{[k]}$.
\item Verify whether or not the matrix $\Phi Z^{[k]}-Y^{[k]}$ is of full column rank.
    \begin{itemize}
    \item If the answer is negative, the dynamic system $\rm\bf\Sigma$ is not observable. End the computations.
    \item If the answer is positive and $k<m$, increase the index $k$ to $k+1$, repeat Steps 2)--4).
    \end{itemize}
\item If $k=m$, the dynamic system $\rm\bf\Sigma$ is observable. End the computations.
\end{enumerate}

Note that the column numbers of both the matrix $Y^{[k]}$ and the matrix $Z^{[k]}$, and therefore that of the matrix $\Phi Z^{[k]}-Y^{[k]}$, are equal to $\sum_{s=1}^{s^{[k]}}p(k,s)$.
From this procedure, it can be seen that if for every $i,j=1,2,\cdots, N$, with $i\neq j$, all the transmission zeros of the TFM $G^{[1]}_{i}(\lambda)$ are different from those of the TFM $G^{[1]}_{j}(\lambda)$, then, the rank condition will be easy to check. Note that under this situation, the number of the columns of these 3  matrices is equal to that of the basis vectors of the null space of {\it only one} subsystem at this transmission zero. But if there are some transmission zeros shared by all the TFMs
$G^{[1]}_{i}(\lambda)|_{i=1}^{N}$, the computational complexity may heavily depends on the connection matrix $\Phi$, due to the fact that the number of the columns of these matrices is equal to the sum of the null space basis vector numbers of {\it all} the plant subsystems at these transmission zeros, which may sometimes be large.

Note that dualities exist between controllability and observability of a dynamic system. That is, controllability of a dynamic system is equivalent to the observability of its dual system, and vice verse \cite{ksh,zdg}. On these basis of these dualities, similar arguments can lead to a computationally efficient algorithm for verifying the controllability of the dynamic system ${\rm\bf\Sigma}$.

More specifically, define TFMs $\bar{G}^{[1]}(\lambda)$, $\bar{G}^{[2]}(\lambda)$, $\bar{G}^{[1]}_{i}(\lambda)$ and $\bar{G}^{[2]}_{i}(\lambda)$, $i=1,2,\cdots,N$, respectively as
\begin{eqnarray*}
& & \bar{G}^{[1]}(\lambda)={\rm\bf diag}\left\{\left.\bar{G}^{[1]}_{i}(\lambda)\right|_{i=1}^{N}\right\},\hspace{0.5cm}
\bar{G}^{[2]}(\lambda)={\rm\bf diag}\left\{\left.\bar{G}^{[2]}_{i}(\lambda)\right|_{i=1}^{N}\right\} \\
& & \bar{G}^{[1]}_{i}(\lambda)=B^{T}_{\rm\bf S}(i)+B^{T}_{\rm\bf T}(i)\left[\lambda I_{m_{\rm\bf Ti}}-A^{T}_{\rm\bf TT}(i)\right]^{-1}A^{T}_{\rm\bf ST}(i), \hspace{0.5cm}
\bar{G}^{[2]}_{i}(\lambda)=\left(G^{[2]}_{i}(\lambda)\right)^{T}
\end{eqnarray*}
Assume that the TFMs $\bar{G}^{[1]}(\lambda)$ has $\bar{m}$ distinctive transmission zeros. Let $\bar{\lambda}_{0}^{[k]}$, $k=1,2,\cdots,\bar{m}$, denote  a transmission zero of the TFM $\bar{G}^{[1]}_{\bar{k}(s)}(\lambda)$, $s=1,2,\cdots,\bar{s}^{[k]}$ with $1\leq \bar{s}^{[k]}\leq N$, $\bar{k}(1)<\bar{k}(2)<\cdots<\bar{k}(\bar{s}^{[k]})$, and $\bar{k}(s)\in \{1,\;2,\;\cdots,\;N\}$. Moreover, let $\bar{\cal Y}^{[k]}_{s}$ denote the set consisting of all complex vectors $\bar{y}_{s,0}^{[k]}$ satisfying $\bar{G}^{[1]}_{\bar{k}(s)}(\bar{\lambda}_{0}^{[k]})\bar{y}_{s,0}^{[k]}=0$. Define $\bar{k}(0)$ as $\bar{k}(0)=0$. On the basis of these vector sets, construct another vector set $\bar{\cal Y}^{[k]}$ as
\begin{equation}
\hspace*{-0.5cm} \bar{\cal Y}^{[k]}\!=\!\left\{y\left|\begin{array}{l}
y\!=\!{\rm\bf col}\!\left\{\!\!\left.\left(0_{m_{{\rm\bf z}(\bar{k}(i)+1)}},\;\cdots,\;0_{m_{{\rm\bf z}(\bar{k}(i+1)-1)}},\; \bar{y}_{i+1,0}^{[k]}\right)\right|_{i=0}^{\bar{s}^{[k]}-1},\;
0_{m_{{\rm\bf z}(\bar{k}(\bar{s}^{[k]})+1)}},\;\cdots,\;0_{m_{{\rm\bf z}N}}\!\right\} \\
\hspace*{2cm} \bar{y}_{i,0}^{[k]}\in \bar{\cal Y}^{[k]}_{i},\;i=1,2,\cdots,\bar{s}^{[k]};\;\; y\neq 0 \end{array}\!\! \right.\!\!\right\}
\end{equation}

Similar to Theorem 2, a necessary and sufficient condition can be derived for the controllability of the dynamic system ${\rm\bf\Sigma}$, which can also lead to a computationally efficient verification procedure. Its proof is given in the appendix.

\hspace*{-0.4cm}{\bf Corollary 1.} The dynamic system ${\rm\bf\Sigma}$ is controllable if and only if for every $k\in\{1,2,\cdots,\bar{m}\}$ and every $\bar{y}^{[k]}\in\bar{\cal Y}^{[k]}$, $\Phi^{T} \bar{G}^{[2]}(\bar{\lambda}_{0}^{[k]})\bar{y}^{[k]}\neq \bar{y}^{[k]}$.

From this corollary, it can be understood that a procedure similar to that of the prototypical observability verification algorithm can also be developed for validating the controllability of the dynamic system ${\rm\bf\Sigma}$, which depends also only on parameter matrices of plant subsystems and the connection matrix. The details are omitted due to their obviousness.

\section{Conditions on the Equivalence of the CDOSSP and the Kalman Filter}

State estimation is a central issue in system science and engineering \cite{kalman,ksh,simon06,zhou10b,zhou11}. When a plant consists of a large number of subsystems or its subsystems are spatially far away from each other, distributed estimations are usually highly appreciated in practical engineering, although their estimation accuracy is generally lower than lumped estimations \cite{kalman,siljak78,simon06}. Recently, a distributed one-step state predictor is suggested in \cite{zhou13} which is generally globally optimal with the widely utilized mean square error criterion under the unbiasedness restriction. Numerical simulations also show that its steady state estimation accuracy may be as high as that of the lumped Kalman filter which has been proved to be optimal for a linear time varying (LTV) system with Gaussian external disturbances and measurement errors. This conclusion, of course, is not valid for every networked dynamic system, noting that the update gain matrix of the Kalman filter is generally not block diagonal.

An interesting problem is therefore that what conditions should be satisfied by a system so that its state prediction accuracy with the distributed algorithm of \cite{zhou13} is equal to that of the lumped Kalman filter. This is investigated in this section using the results of Section III.

As in \cite{zhou13}, this state predictor is called CDOSSP for brevity in the following discussions, which is an abbreviation for Coordinated Distributed One-step State Predictor.

\subsection{A brief summary of the distributed prediction algorithm and some preliminaries}

In developing a distributed state prediction procedure, a system representation similar to that of Equations (\ref{eqn:7}) and (\ref{eqn:8}) is adopted in \cite{zhou13}. More precisely,  concerning a networked system $\bar{\rm\bf\Sigma}$
consisting of $N$ LTV dynamic subsystems, assume the dynamics of its $i$-th
subsystem ${\bar{\rm\bf\Sigma}}_{i}$ and influences among its subsystems are respectively described by
the following discrete state-space model and equality,
\begin{eqnarray}
& & \left[\!\!\begin{array}{c} x(t\!+\!1,i) \\ z(t,i) \\
y(t,i)
\end{array}\!\!\right] \!\!=\!\!\left[\!\!\begin{array}{ccc} A_{\rm\bf TT}(t,i) & A_{\rm\bf TS}(t,i) & B_{\rm\bf T}(t,i)  \\
A_{\rm\bf ST}(t,i) & A_{\rm\bf SS}(t,i) & 0 \\
C_{\rm\bf T}(t,i) & C_{\rm\bf S}(t,i) & 0
\end{array}\!\!\right.\! \left.\!\!\begin{array}{c} 0 \\  0 \\
 D_{\rm\bf w}(t,i)
\end{array}\!\!\right]\! \left[\!\!\begin{array}{c} x(t,i)
\\ v(t,i) \\ d(t,i) \\ w(t,i)
\end{array}\!\!\right]  \label{eqn:1} \\
& &  v(t)=\Phi(t) z(t) \label{eqn:2}
\end{eqnarray}
Here, the vectors $x(t,i)$, $y(t,i)$, $v(t,i)$, etc., as well as the matrices $A_{\rm\bf TT}(t,i)$, $A_{\rm\bf TS}(t,i)$, $C_{\rm\bf T}(t,i)$, etc., have completely the same meanings as their counterparts of the Equations (\ref{eqn:7}) and (\ref{eqn:8}).

In the following discussions, all matrices like $A_{\rm\bf SS}(t)$, $B_{\rm\bf T}(t)$, $A(t)$, etc., are defined in completely the same way as their time invariant counterparts.

The structure and dynamics of the above system are very similar to those of the system $\rm\bf\Sigma$ described by Equations (\ref{eqn:7}) and (\ref{eqn:8}). However, the system matrices of every subsystem in the system $\bar{\rm\bf\Sigma}$ are permitted to be time dependent, as well as the connection matrix among its subsystems, which makes this model capable of describing dynamics of a wider class of plants. For example, a plant with time varying topology, etc. On the other hand, direct pass from external input vector $d(t,i)$ to both the internal output vector $z(t,i)$ and the external output vector $y(t,i)$ are not permitted, which is reflected by that both the matrices $B_{\rm\bf S}(i)$ and $D_{\rm\bf d}(i)$ of Equation (\ref{eqn:7}) are replaced by a zero matrix. This is only for avoiding complicated mathematical expressions and awkward statements, noting that in estimating problems, correlations between process disturbances and output measurements usually lead to awkward equations. A widely adopted method to deal with this situation is to introduce some transformations that decouple this correlation \cite{ksh}.

In \cite{zhou13}, it is suggested to predict the system's states using an observer that has the same structure as that of the plant. More specifically, the observer also consists of $N$
subsystems, and the input-output relation of its $i$-th subsystem $\hat{\rm\bf\Sigma}_{i}$ and interactions among its subsystems are respectively described as
\begin{eqnarray}
& & \left[\!\!\begin{array}{c} \hat{x}(t\!+\!1,i) \\ \hat{z}(t,i) \\
\hat{y}(t,i)
\end{array}\!\!\right] \!\!=\!\!\left[\!\!\begin{array}{ccc} A_{\rm\bf TT}(t,i) & A_{\rm\bf TS}(t,i) & K_{\rm\bf T}(t,i)  \\
A_{\rm\bf ST}(t,i) & A_{\rm\bf SS}(t,i) & 0 \\
C_{\rm\bf T}(t,i) & C_{\rm\bf S}(t,i) & 0
\end{array}\!\!\right]\left[\!\!\begin{array}{c} \hat{x}(t,i)
\\ \hat{v}(t,i) \\ y(t,i)-\hat{y}(t,i)
\end{array}\!\!\right]  \label{eqn:3} \\
& & \hat{v}(t)=\Phi(t) \hat{z}(t) \label{eqn:4}
\end{eqnarray}
Here, $\hat{z}(t)={\rm\bf col}\!\left\{\hat{z}(t,i)|_{i=1}^{N}\right\}$
and $\hat{v}(t)={\rm\bf
col}\!\left\{\hat{v}(t,i)|_{i=1}^{N}\right\}$. Clearly, in estimating the states of a subsystem, only local output measurements are utilized. This is very attractive in implementing state estimations in a distributed way for a large scale networked system or a system having geographically widely spread subsystems, etc.

Under the criteria of unbiasedness and estimation error variance minimization, it is proved in \cite{zhou13} that when the external input vectors $d(t,i)$ and $w(t,i)$ are white and uncorrelated random processes, both the optimal observer gain matrices $K_{\rm\bf T}(t,i)|_{i=1}^{N}$ and the covariance matrix of estimation errors can be recursively computed. An interesting property of the suggested estimation procedure is that in these computations, the dimensions of the involved matrices depend {\it only} on the dimensions of every subsystem, which makes it attractive in state estimations for a networked system with a great amount of states.

More precisely, let $\tilde{x}(t)$ and $\tilde{x}(t,i)$ denote state prediction errors respectively for the whole system and its $i$-th subsystem, and
denote ${\rm\bf E}\{\tilde{x}(t)\tilde{x}^{T}(t)\}$ and ${\rm\bf E}\{\tilde{x}(t,i)\tilde{x}^{T}(t,j)\}$ respectively by $P(t)$ and
$P_{ij}(t)$, $i,j=1,2,\cdots,N$. Let $J_{{\rm\bf T}i}$ and $J_{{\rm\bf S}i}$ represent
respectively matrices ${\bf col}\!\left\{\!0_{M_{{\rm\bf T}i}\times
m_{{\rm\bf T}i}}, I_{m_{{\rm\bf T}i}}, 0_{(M_{\rm\bf T}-M_{{\rm\bf
T}\!,i+1})\times m_{{\rm\bf T}i}}\!\right\}$ and ${\bf
col}\!\left\{\! 0_{M_{{\rm\bf S}i}\times m_{{\rm\bf S}i}},
I_{m_{{\rm\bf S}i}}, 0_{(M_{\rm\bf S}-M_{{\rm\bf S}\!,i+1})\times
m_{{\rm\bf S}i}}\!\right\}$. Moreover, let $A_{\rm\bf T}(t,i)$ and
$C(t,i)$ stand respectively for matrices $[A_{\rm\bf TT}(t,i) \;\; A_{\rm\bf
TS}(t,i)]$ and $[C_{\rm\bf T}(t,i) \;\;C_{\rm\bf S}(t,i)]$, and define
matrix $W(t,i)$ as
\begin{displaymath}
W(t,i)=\left[\begin{array}{cc} J_{{\rm\bf T}i}^{T} \\
J_{{\rm\bf S}i}^{T}\Phi(t)\left[\;I-A_{\rm\bf
SS}(t)\Phi(t)\;\right]^{-1}\! A_{\rm\bf ST}(t)\end{array}\right]
\end{displaymath}
Then, under the condition that the matrix $D_{\rm\bf w}(t,i)$ is of full row rank for every $i=1,2,\cdots,N$, it has been proved in \cite{zhou13} that for every
subsystem $\bar{\rm\bf \Sigma}_{i}$ and every time instant $t$, the
optimal gain matrix $K_{\rm\bf T}(t,i)$, which is denoted by $K_{\rm\bf T}^{opt}(t,i)$, can be expressed as
\begin{eqnarray}
& & \hspace*{-0.5cm}K_{\rm\bf T}^{opt}(t,i)\!=\!A_{\rm\bf
T}(t,i)\left\{I+W(t,i)P(t)W^{T}(t,i)C^{T}(t,i)\left[D_{\rm\bf w}(t,i)D^{T}_{\rm\bf w}\!(t,i)\right]^{-1}\!\!C(t,i)\!
\right\}^{\!-1}\!\times\nonumber\\
& &
\hspace*{6.4cm}W(t,i)P(t)W^{T}(t,i)C^{T}(t,i)\left[D_{\rm\bf w}(t,i)D_{\rm\bf w}^{T}(t,i)\right]^{-1}
\label{eqn:5}
\end{eqnarray}
Moreover, for every $i,j=1,2,\cdots,N$, the $i$-th row $j$-th column block
matrix of the covariance matrix $P(t+1)$ can be recursively
expressed as follows,
\begin{equation}
\hspace*{-0.0cm}P_{ij}(t\!+\!1)\!\!=\!\!\left\{\!\!\!\!\begin{array}{ll}
A_{\rm\bf
T}(t,i)\left\{I+W(t,i)P(t)W^{T}(t,i)C^{T}(t,i)\left[D_{\rm\bf w}(t,i)D^{T}_{\rm\bf w}\!(t,i)\right]^{-1}\!\!C(t,i)\!
\right\}^{\!-1}\!\!\!\!\!\times  & \\
\hspace*{3.25cm}W(t,i)P(t)W^{T}(t,i)A_{\rm\bf
T}^{T}(t,i)\!+\!B_{\rm\bf T}(t,i)B_{\rm\bf
T}^{T}(t,i) &  i=j \\
A_{\rm\bf
T}(t,i)\left\{I\!+\!W(t,i)P(t)W^{T}(t,i)C^{T}(t,i)\left[D_{\rm\bf w}(t,i)D^{T}_{\rm\bf w}\!(t,i)\right]^{-1}\!\!C(t,i)\!
\right\}^{-1}\!\!\!W(t,i)P(t)\!\times & \\
\hspace*{0.25cm}W^{T}(t,j)\left\{\!\!I\!+\!C^{T}(t,j)\left[D_{\rm\bf w}(t,j)D^{T}_{\rm\bf w}\!(t,j)\right]^{\!-1}\!C(t,j)W(t,\!j)P(t)W^{T}(t,\!j)\!
\right\}^{-1}\!\!\!A_{\rm\bf T}^{T}(t,j) & i\neq j \\
\end{array}\right.
\label{eqn:6}
\end{equation}

\begin{figure}[!ht]
\begin{center}
\includegraphics[width=3.8in]{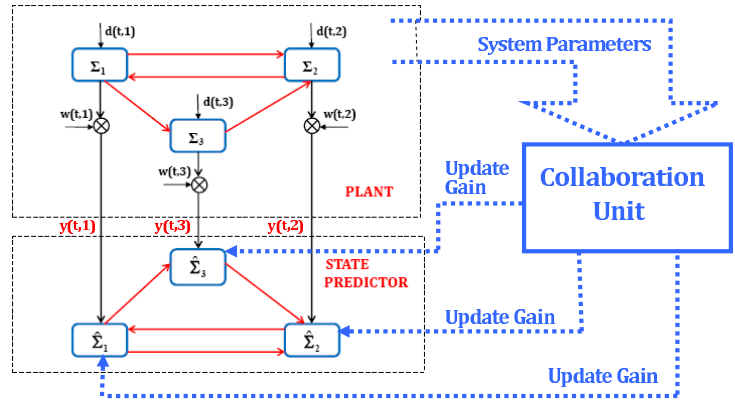}
\end{center}

\vspace{-0.8cm} \caption{Schematic Diagram for the Implementation of the
CDOSSP for a Plant with 3 Subsystems}
\end{figure}

\vspace{-0.5cm} Figure 1 gives a schematic diagram for the implementation of this one-step
state predictor for a plant with $3$ subsystems, together with the plant itself.

In \cite{zhou13}, it is proved that for every subsystem, prediction accuracy of this distributed state predictor is generally worse than that of the lumped Kalman filter. This is not a surprising result, as in estimating states of each subsystem, the above predictor utilizes only the output measurement of the subsystem itself, that is, it utilizes less information than the lumped Kalman filter. However, numerical simulations in that paper show that for some systems, the CDOSSP and the Kalman filter may have almost the same steady state estimation accuracy, even through their subsystems are coupled. This is an interesting observation, as it means that even for a system with coupled subsystems, a distributed estimator may sometimes perform as good as a lumped one.

Note that it is declared in \cite{zhou13} that this estimation procedure can be extended to situations under which some of the matrices $D_{\rm\bf w}(t,i)|_{i=1}^{N}$ are not of full row rank. However, our attention in this paper is concentrated on the case when every $D_{\rm\bf w}(t,i)$ is left invertible. This is only for avoiding awkward statements and complicated mathematical expressions.

To clarify the aforementioned situations, some necessary and sufficient conditions are given in the remaining of this section on system matrices such that this  equivalence is theoretically guaranteed.

The next lemma is concerned with convergence properties of Kalman filtering, which is now widely known in estimation theories \cite{simon06,ksh}.

\hspace*{-0.4cm}{\bf Lemma 4.} Concerning plants having a state space representation of Equation (\ref{eqn:9}), assume that ${\rm\bf col}\{d(t),\;$ $w(t)\}$ are white Gaussian stationary random process with a zero mean and an identity covariance matrix. Moreover, assume that for every complex scalar $\lambda$ and every nonzero complex vector $x$ satisfying $x^{H}A=\lambda x^{H}$ and $|\lambda|=1$, $x^{H}B\neq 0$. Furthermore, assume that for every complex scalar $\lambda$ and every nonzero complex vector $y$ satisfying $Ay=\lambda y$ and $|\lambda|\geq 1$, $Cy\neq 0$. Then, from every semi-positive definite initial covariance matrix of estimation errors, with the increment of the temporal variable $t$, both the covariance matrix of estimation errors and the update gain matrix of its Kalman filter converge respectively to the unique stabilizing solution of a discrete algebraic Riccati equation (DARE) and a constant matrix.

It is worthwhile to mention that while in the above lemma, only a sufficient condition is given for the convergence of the Kalman filter, it is widely believed that this condition is also necessary, as this condition is equivalent to the detectability and the unit circle stability of the plant, which is necessary for the boundedness of the covariance matrix of estimation errors \cite{simon06,ksh,zhou10b,zhou11}.

\subsection{Conditions on the equivalence}

From Lemma 4, it is clear that convergence of the Kalman filter is closely related to the controllability and observability of the plant to be estimated. As a matter of fact, the conditions of Lemma 4 are equal to system controllability with every mode on the unit circle and system observability with every unstable mode \cite{ksh,zdg}. Moreover, it can be directly declared from Lemmas 1 and 4 that if a LTI system is both controllable and observable, then, both the update gain matrix of its Kalman filter and the covariance matrix of the corresponding estimation errors will certainly converge to constant matrices.

When steady state behaviors of an estimation algorithm is to be investigated, the plant is usually assumed to be time invariant, while the external inputs stationary. Under the time invariance condition, the dynamic system $\bar{\rm\bf\Sigma}$ is completely equal to the dynamic system $\rm\bf\Sigma$ with $B_{\rm\bf S}(i)\equiv 0$ and $D_{\rm\bf d}(i)\equiv 0$.

The above arguments mean that results of Section III are helpful in the convergence analysis of the CDOSSP, which is made clear by the following Corollary 2.

Define the sets $\rm\bf\Lambda$ and $\bar{\rm\bf\Lambda}$ respectively as
\begin{equation}
{\rm\bf\Lambda}=
\left\{\lambda_{0}^{[1]},\;\lambda_{0}^{[2]},\;\cdots,\; \lambda_{0}^{[m]}\right\},\hspace{0.5cm}
\bar{\rm\bf\Lambda}= \left\{\bar{\lambda}_{0}^{[1]},\;\bar{\lambda}_{0}^{[2]},\;\cdots,\; \bar{\lambda}_{0}^{[\bar{m}]}\right\}
\end{equation}
Then, on the basis of Theorem 2 and Corollary 1, an "almost" necessary and sufficient condition can be established for the Kalman filter when the dynamics of a plant is described by Equations (\ref{eqn:7}) and (\ref{eqn:8}) with $B_{\rm\bf S}(i)\equiv 0$ and $D_{\rm\bf d}(i)\equiv 0$.

\hspace*{-0.4cm}{\bf Corollary 2.} Assume that for each $i=1,2,\cdots,N$, $B_{\rm\bf S}(i)=0$ and $D_{\rm\bf d}(i)=0$. Then, the Kalman filter of the dynamic system $\rm\bf\Sigma$ converges to a LTI observer, if for every $\lambda_{0}^{[k]}\in {\rm\bf\Lambda}$ with $|\lambda_{0}^{[k]}|\geq 1$,  $\Phi G^{[2]}(\lambda_{0}^{[k]})y^{[k]}\neq y^{[k]}$
with each $y^{[k]}\in{\cal Y}^{[k]}$, as well as
for every $\bar{\lambda}_{0}^{[k]}\in \bar{\rm\bf\Lambda}$ with $|\bar{\lambda}_{0}^{[\bar{k}]}|=1$,  $\Phi^{T} \bar{G}^{[2]}(\bar{\lambda}_{0}^{[\bar{k}]})\bar{y}^{[\bar{k}]}\neq \bar{y}^{[\bar{k}]}$
with each $\bar{y}^{[\bar{k}]}\in\bar{\cal Y}^{[\bar{k}]}$.

The proof of Corollary 2 is deferred to the appendix.

Note that a transmission zero of the TFM $G^{[1]}(\lambda)$ is not necessary an eigenvalue of the matrix $A$. The results of Corollary 2 are to some extent surprising, as their relations with Theorem 2 and Corollary 1 are completely the same as those between Lemmas 1 and 4.

From Corollary 2, it can be claimed that if the TFM $G^{[1]}(\lambda)$ does not have a transmission zero with a magnitude greater than $1$ and the TFM $\bar{G}^{[1]}(\lambda)$ does not have a transmission zero belonging to the unit circle, then, the Kalman filter of the networked system $\rm\bf\Sigma$ will certain converge.

Corollary 2 also clearly suggests that when only convergence properties of Kalman filtering are to be investigated for the dynamic system $\rm\bf\Sigma$, similar but usually less computations are to be implemented than verifying controllability and observability of this dynamic system, as the required conditions are to be verified in general only for part of the elements in the set $\rm\bf\Lambda$ and $\bar{\rm\bf\Lambda}$.

To establish conditions under which estimation accuracies in the steady states of the CDOSSP are equivalent to those of the Kalman filter, we need to investigate differences in the covariance matrices between these two predictors.

The following lemma gives another expression for the covariance matrix $P(t+1)$ which will be used in the following mathematical arguments. Its proof is given in \cite{zhou13} under the title Theorem 3.

\hspace*{-0.4cm}{\bf Lemma 5.} Assume that the matrix $D_{\rm\bf w}(t)$ is of
full row rank. Define matrices $J_{{\rm\bf y}i}$, $\bar{C}_{i}(t)$ and $A_{i}(t)$
respectively as $J_{{\rm\bf y}i}={\bf col}\!\left\{\!0_{M_{{\rm\bf y}i}\times
m_{{\rm\bf y}i}}, I_{m_{{\rm\bf y}i}}, 0_{(M_{\rm\bf y}-M_{{\rm\bf
y}\!,i+1})\times m_{{\rm\bf y}i}}\!\right\}$, $\bar{C}_{i}(t)=J_{{\rm\bf
y}i}^{T}[D(t)D^{T}(t)]^{-1/2}C(t)$ and $A_{i}(t)=J_{{\rm\bf
T}i}^{T}A(t)$, $i=1,2,\cdots,N$. Then, for each $i,j=1,2,\cdots,N$, the covariance matrix
$P_{ij}(t+1)$ has the following equivalent expression,
\begin{equation}
P_{ij}(t+1)=\left\{\begin{array}{ll}
A_{i}(t)\left[P^{-1}(t)+\bar{C}_{i}^{T}(t)\bar{C}_{i}(t)\right]^{-1}\!\!A_{i}^{T}(t)+B_{\rm\bf
T}(t,i)B_{\rm\bf T}^{T}(t,i) & i=j \\
A_{i}(t)\left[P^{-1}(t)+\bar{C}_{i}^{T}(t)\bar{C}_{i}(t)\right]^{-1}\!\!P^{-1}(t)
\left[P^{-1}(t)+\bar{C}_{j}^{T}(t)\bar{C}_{j}(t)\right]^{-1}\!\!A_{j}^{T}(t) &
i\neq j
\end{array}\right.
\label{eqn:16}
\end{equation}

On the basis of these results, differences can be established between the covariance matrices of estimation errors for each subsystem ${\bar{\rm\bf\Sigma}}_{i}$ of the Kalman filter and the CDOSSP. These results are given in \cite{zhou13} without proof. Due to their importance in the following analyses, a proof is given in the appendix.

\hspace*{-0.4cm}{\bf Lemma 6.} Let $P^{[kal]}(t)$ denote the covariance matrix when the lumped Kalman filter is applied to the dynamic system $\bar{\rm\bf\Sigma}$, and $P_{ij}^{[kal]}(t)$ its $i$-th row $j$-th column block. Assume that the matrix $D_{\rm\bf w}(t)$ is of full row rank. Then,
\begin{eqnarray}
& &
P_{ii}(t+1)-{P}_{ii}^{[kal]}(t+1)=A_{i}(t)\left\{[{P}^{[kal]}(t)]^{-1}+\sum_{k=1}^{N}\bar{C}_{k}^{T}(t)\bar{C}_{k}(t)\right\}^{-1}\times\nonumber\\
& & \hspace*{1cm}
\left\{[{P}^{[kal]}(t)]^{-1}-P^{-1}(t)+\sum_{k=1,k\neq
i}^{N}\!\!\!\bar{C}_{k}^{T}(t)\bar{C}_{k}(t)\right\}[P^{-1}(t)+\bar{C}_{i}^{T}(t)\bar{C}_{i}(t)]^{-1}A_{i}^{T}(t)
\end{eqnarray}

Based on Lemma 6, a necessary and sufficient condition can be derived on system matrices for the equivalence between the estimation accuracy of the CDOSSP and that of the Kalman filter.

\hspace*{-0.4cm}{\bf Theorem 3.} Assume that $P(t)=P^{[kal]}(t)$ at some time instant $t$. Then, at the next time instant $t+1$, for the $i$-th plant subsystem ${\bar{\rm\bf\Sigma}}_{i}$, the CDOSSP has a covariance matrix of estimation errors equal to that of the Kalman filter, if and only if for each  $j=1,2,\cdots,N$, $j\neq i$,
\begin{equation}
A_{i}(t)P(t)\bar{C}^{T}(t)[I+\bar{C}(t)P(t)\bar{C}^{T}(t)]^{-1}J_{{\rm\bf
y}\!j}=0
\end{equation}
in which $\bar{C}(t)={\rm\bf col}\{\bar{C}_{i}(t)|_{i=1}^{N}\}$. Moreover, let $K^{[kal]}(t)$ denote the update gain matrix of the Kalman filter and $K_{ij}^{[kal]}(t)$ its $i$-th row $j$-th column block. Then, when this condition is satisfied, the next two equalities are also valid.
\begin{equation}
P_{ij}(t+1)=P_{ij}^{[kal]}(t+1),\hspace{0.25cm}
K_{ij}^{[kal]}(t)=0,\hspace{0.5cm}\forall j\neq i
\end{equation}

The proof of Theorem 3 is deferred to the appendix.

Note that in the steady state of an estimation procedure, the covariance matrix of its estimation errors keeps unchanged.
On the basis of this observation and Theorem 3, systems are characterized whose steady state estimation accuracy with the CDOSSP is equal to that of the Kalman filter.
The results are given in the next theorem, while their proof is provided in the appendix.

\hspace*{-0.4cm}{\bf Theorem 4.} Assume that both the subsystem parameter matrices of the system $\bar{\rm\bf\Sigma}$ and its subsystem connection matrix are time independent. Moreover, assume that the covariance matrix of estimation errors of its Kalman filter converges to a constant and positive definite matrix $P$. Then, the CDOSSP can have the same steady state estimation accuracy as the Kalman filter, if and only if for every $i=1,2,\cdots,N$,
\begin{equation}
A_{i}P\bar{C}^{T}[I+\bar{C}P\bar{C}^{T}]^{-1}J_{{\rm\bf
y}\!j}=0,\hspace{0.5cm}\forall j\neq i
\label{eqn:40}
\end{equation}

From Theorems 3 and 4, it is clear that in order to guarantee that the CDOSSP has the same estimation accuracy in its steady state as the Kalman filter, it is necessary that the Kalman filter has a block diagonal update gain matrix in its steady state for a system whose states are to be predicted.

Theorem 4, together with Corollary 2, gives almost necessary and sufficient conditions for a system whose steady state covariance matrix with the CDOSSP is equal to that of the Kalman filter. While these results are helpful in understanding properties of the CDOSSP and can be utilized in analyzing and synthesizing a networked system with some subsystems, they are still difficult to be applied to a large scale system, especially those with a great amount of states. It is because that in the conditions of Theorem 4, both the system matrices of the whole networked system and the steady state covariance matrix of its Kalman filter are required, and either of them is usually very difficult to be obtained when the number of subsystems is large and/or the connection matrix is of a high dimension.

On the other hand, in the proof of Theorem 4, convergence of the CDOSSP is established on the condition of initializing it with the steady state covariance matrix of the Kalman filter. It is interesting to see whether or not these results can be extended to other situations.

\section{Concluding Remarks}

This paper investigates controllability and observability of a networked system with LTI (linear time invariant) subsystems and time independent subsystem connections. The plant subsystems can have different TFMs (transfer function matrix) and their connections can be arbitrary. Some necessary and sufficient conditions have been derived on the basis of the PBH test. These conditions only depend on the subsystem connection matrix and parameter  matrices of the plant subsystems, and have some nice physical interpretations. This characteristic makes these conditions easily implementable in general on a large scale networked system.

Convergence of the Kalman filter for a networked system has also been discussed. Conditions similar to those for controllability and observability are established. Based on these results, networked systems are characterized that have a steady state covariance matrix of estimation errors with the state predictor suggested in \cite{zhou13} equal to that of the Kalman filter. It has been clarified that to guarantee this equivalence, the update gain matrix of the Kalman filter must be block diagonal.

It is interesting to see whether or not these results can be extended to networked systems with random communication delays and data missing, etc. Challenging issues also include computationally more attractive conditions on system matrices for the equivalence in estimation accuracy between the lumped Kalman filter and the CDOSSP.

\renewcommand{\theequation}{a.\arabic{equation}}
\setcounter{equation}{0}

\section*{Appendix: Proof of the Technical Results}

\hspace*{-0.4cm}{\bf Proof of Theorem 1:} Assume that the LTI dynamic  system ${\rm\bf\Sigma}$ is observable. Then, according to Lemma 1, for every scalar complex number $\lambda$ and every nonzero $M_{\rm\bf T}$ dimensional complex vector $y$ satisfying $(\lambda I_{M_{\rm\bf T}}-A)y=0$, $Cy\neq 0$. From the definitions of the matrices $A$ and $C$, it is clear that this is equivalent to that there does not exist a pair of complex scalar $\lambda$ and complex vector $y$ with $y\neq 0$ such that
\begin{equation}
\left[\begin{array}{c} \lambda I_{M_{\rm\bf T}}-\left[A_{\rm\bf TT}+A_{\rm\bf TS}
(I_{M_{\rm\bf S}}-\Phi A_{\rm\bf SS})^{-1}\Phi A_{\rm\bf ST}\right] \\
C_{\rm\bf T}+C_{\rm\bf S}
(I_{M_{\rm\bf S}}-\Phi A_{\rm\bf SS})^{-1}\Phi A_{\rm\bf ST}\end{array}\right]y=0 \label{eqn:a6}
\end{equation}

Now, assume that there exists a complex scalar $\lambda$, denote it by $\lambda_{0}$, such that the MVP $M(\lambda)$ is not of full column rank. Then, there must exists a nonzero $M_{\rm\bf T}+M_{\rm\bf S}$ dimensional complex vector $z$ such that
\begin{equation}
M(\lambda_{0})z=0 \label{eqn:a1}
\end{equation}

Partition the vector $z$ as $z={\rm\bf col}\{z_{1},\;z_{2}\}$ with $z_{1}$ and $z_{2}$ respectively having a dimension of $M_{\rm\bf T}$ and $M_{\rm\bf S}$. Then, according to Equation (\ref{eqn:a1}) and the definition of $M(\lambda)$, we have
\begin{eqnarray}
& & (\lambda_{0} I_{M_{\rm\bf T}}-A_{\rm\bf TT})z_{1}-A_{\rm\bf TS}z_{2}=0 \label{eqn:a2}\\
& & C_{\rm\bf T}z_{1}+C_{\rm\bf S}z_{2}=0 \label{eqn:a3} \\
& & -\Phi A_{\rm\bf ST}z_{1}+(I_{M_{\rm\bf S}}-\Phi A_{\rm\bf SS})z_{2}=0
\label{eqn:a4}
\end{eqnarray}
From the state space representation like model of System ${\rm\bf\Sigma}$ given by  Equations (\ref{eqn:7}) and (\ref{eqn:8}) and its well-posedness, it can be directly proved that the matrix $I_{M_{\rm\bf S}}-\Phi A_{\rm\bf SS}$ is invertible \cite{zhou13}. It can therefore be claimed from Equation (\ref{eqn:a4}) that $z_{1}\neq 0$ and
\begin{equation}
z_{2}=(I_{M_{\rm\bf S}}-\Phi A_{\rm\bf SS})^{-1}\Phi A_{\rm\bf ST}z_{1}
\label{eqn:a5}
\end{equation}

Substitute this expression for the vector $z_{2}$ respectively into Equations (\ref{eqn:a2}) and (\ref{eqn:a3}), direct algebraic manipulations show that
\begin{eqnarray}
& & \left\{\lambda_{0} I_{M_{\rm\bf T}}-\left[A_{\rm\bf TT}+A_{\rm\bf TS}
(I_{M_{\rm\bf S}}-\Phi A_{\rm\bf SS})^{-1}\Phi A_{\rm\bf ST}\right]\right\}z_{1}=0 \\
& & \left[C_{\rm\bf T}+C_{\rm\bf S}
(I_{M_{\rm\bf S}}-\Phi A_{\rm\bf SS})^{-1}\Phi A_{\rm\bf ST}\right]z_{1}=0
\end{eqnarray}
These two equations and Equation (\ref{eqn:a6}) clearly contradict each other, which means that the existence of the aforementioned $\lambda_{0}$ and $z$ is not possible, and therefore the MVP $M(\lambda)$ is of full column rank for every complex number $\lambda$.

On the contrary, assume that the MVP $M(\lambda)$ is of full column rank for every complex scalar $\lambda$, but the dynamic system ${\rm\bf\Sigma}$ is not observable. Then, according to Lemma 1 and the definitions of the matrices $A$ and $C$, there exist at least one scalar complex number $\lambda$ and one nonzero $M_{\rm\bf T}$ dimensional complex vector $y$ such that the following two equalities are simultaneously satisfied.
\begin{eqnarray}
& & \left\{\lambda I_{M_{\rm\bf T}}-\left[A_{\rm\bf TT}+A_{\rm\bf TS}
(I_{M_{\rm\bf S}}-\Phi A_{\rm\bf SS})^{-1}\Phi A_{\rm\bf ST}\right]\right\}y=0 \label{eqn:a7}\\
& & \left[C_{\rm\bf T}+C_{\rm\bf S}
(I_{M_{\rm\bf S}}-\Phi A_{\rm\bf SS})^{-1}\Phi A_{\rm\bf ST}\right]y=0
\label{eqn:a8}
\end{eqnarray}

Define vector $\psi$ as $\psi=(I_{M_{\rm\bf S}}-\Phi A_{\rm\bf SS})^{-1}\Phi A_{\rm\bf ST} y$. Then, it can be straightforwardly proved from Equations (\ref{eqn:a7}) and (\ref{eqn:a8}), as well as the definition of the vector $\psi$, that for this complex scalar $\lambda$,
\begin{equation}
M(\lambda)\left[\begin{array}{c} y \\ \psi \end{array}\right]=0 \label{eqn:a9}
\end{equation}
As the vector $y$ is not equal to zero, it is clear that ${\rm\bf col}\{y,\; \psi\}$ is also not a zero vector. This means that Equation (\ref{eqn:a9}) is a contradiction to the assumption on the MVP  $M(\lambda)$. Hence, the dynamic system ${\rm\bf\Sigma}$ must be observable.

This completes the proof.  \hspace{\fill}$\Diamond$

\hspace*{-0.4cm}{\bf Proof of Lemma 3:} Assume that the MVP $M(\lambda)$ defined in Equation (\ref{eqn:14}) is not always of full column rank. Then, there exist at least one complex scalar $\lambda_{0}$ and a corresponding nonzero complex vector $z$, such that
\begin{equation}
M(\lambda_{0})z=0
\label{eqn:a10}
\end{equation}

Partition the vector $z$ as $z={\rm\bf col}\{z_{1},\;z_{2}\}$, in which $z_{1}$ is a $M_{\rm\bf T}$ dimensional complex vector, while $z_{2}$ has a compatible dimension. Then, from Equation (\ref{eqn:a10}) we have that $\left(\lambda_{0} I_{M_{\rm\bf T}}-A_{\rm\bf TT}\right)z_{1}-A_{\rm\bf TS}z_{2}=0$. As the matrix $A_{\rm\bf TT}$ is square according to its definition, it is clear that the matrix $\lambda I_{M_{\rm\bf T}}-A_{\rm\bf TT}$ is of full normal rank. Hence, $z_{1}$ can always be formally expressed as\footnote[1]{If the matrix $\lambda_{0} I_{M_{\rm\bf T}}-A_{\rm\bf TT}$ is not invertible at some particular $\lambda_{0}$ which is usually possible, $A_{\rm\bf TS}z_{2}$ must belong to the space spanned by the vectors of $\lambda_{0} I_{M_{\rm\bf T}}-A_{\rm\bf TT}$. This guarantees the validness of the adopted expression for $z_{2}$ with the matrix inverse being interpreted as the generalized inverse \cite{hj91}.}
\begin{equation}
z_{1}=\left(\lambda_{0} I_{M_{\rm\bf T}}-A_{\rm\bf TT}\right)^{-1}A_{\rm\bf TS}z_{2}
\label{eqn:a11}
\end{equation}

Substitute this relation between $z_{1}$ and $z_{2}$ back into Equation (\ref{eqn:a10}), straightforward algebraic manipulations show that
\begin{eqnarray}
& & \left[C_{\rm\bf S}+C_{\rm\bf T}\left(\lambda_{0} I_{M_{\rm\bf T}}-A_{\rm\bf TT}\right)^{-1}A_{\rm\bf TS}\right]z_{2}=0 \\
& & \left[\Phi A_{\rm\bf SS}-I_{M_{\rm\bf S}}+\Phi A_{\rm\bf ST}\left(\lambda_{0} I_{M_{\rm\bf T}}-A_{\rm\bf TT}\right)^{-1}A_{\rm\bf TS}\right]z_{2}=0
\end{eqnarray}
From the definition of the TFM $G(\lambda)$, it is clear that simultaneous satisfaction of these two equalities is equivalent to $G(\lambda_{0})z_{2}=0$. As the vector $z$ is not equal to zero, Equation (\ref{eqn:a11}) and $z={\rm\bf col}\{z_{1},\;z_{2}\}$ also imply that $z_{2}$ is not a zero vector either. Hence, according to Lemma 2, $\lambda_{0}$ is a transmission zero of the TFM $G(\lambda)$.

On the contrary, assume that the TFM $G(\lambda)$ has transmission zeros and let $\lambda_{0}$ denote one of them. According to Lemma 2, there exists a nonzero vector $z_{0}$ such that $G(\lambda_{0})z_{0}=0$. Define a vector $y_{0}$ as $y_{0}=\left(\lambda_{0} I_{M_{\rm\bf T}}-A_{\rm\bf TT}\right)^{-1}A_{\rm\bf TS}z_{0}$. Then, from the definition of the TFM $G(\lambda)$, it can be directly proved that
\begin{equation}
M(\lambda_{0})\left[\begin{array}{c} z_{0} \\ y_{0} \end{array}\right]=0
\end{equation}
As $z_{0}$ is not a zero vector, it is clear that ${\rm\bf col}\{z_{0},\;y_{0}\}$ is not a zero vector too. This further means that the MVP polynomial $M(\lambda)$ is not of full column rank at $\lambda_{0}$.

This completes the proof.  \hspace{\fill}$\Diamond$

\hspace*{-0.4cm}{\bf Proof of Theorem 2:} From the definitions of matrices $A_{\rm\bf SS}$, $A_{\rm\bf ST}$, $A_{\rm\bf TS}$, $A_{\rm\bf TT}$, $C_{\rm\bf S}$ and $C_{\rm\bf T}$, it is clear that all of them are block diagonal with consistent dimensions and their block diagonal matrices are in fact system matrices of the plant subsystems. On the basis of this observation, and the definitions of the TFMs $G(\lambda)$, $G^{[1]}(\lambda)$ and $G^{[2]}(\lambda)$, straightforward algebraic manipulations show that
\begin{equation}
G(\lambda)=\left[\begin{array}{c} G^{[1]}(\lambda) \\
\Phi G^{[2]}(\lambda)-I_{M_{\rm\bf S}} \end{array}\right]
\label{eqn:a12}
\end{equation}

Assume that the dynamic system ${\rm\bf \Sigma}$ is observable, but there exists an integer $k$ belonging to the set $\{1,2,\cdots,m\}$ and a complex vector $y^{[k]}$ belonging to the set ${\cal Y}^{[k]}$, such that $\Phi G^{[2]}(\lambda_{0}^{[k]})y^{[k]}=y^{[k]}$. Then, according to the definition of the set ${\cal Y}^{[k]}$, we have that $y^{[k]}\neq 0$ and
$G^{[1]}(\lambda_{0}^{[k]})y^{[k]}=0$. It can therefore be declared from Equation (\ref{eqn:a12}) that
\begin{equation}
G(\lambda_{0}^{[k]})y^{[k]}=\left[\begin{array}{c} G^{[1]}(\lambda_{0}^{[k]})y^{[k]} \\
\Phi G^{[2]}(\lambda_{0}^{[k]})y^{[k]}-y^{[k]} \end{array}\right]=0
\end{equation}
 Based on Lemma 2, it is clear that the complex number $\lambda_{0}^{[k]}$ is a transmission zero of the TFM $G(\lambda)$. Then, it can be further claimed from Lemma 3 and Theorem 1 that the dynamic system ${\rm\bf \Sigma}$ is not observable, which is a contradiction to the observability assumption on this dynamic system. Therefore, the assumption about the existence of $1\leq k\leq m$ and $y^{[k]}\in {\cal Y}^{[k]}$ satisfying $\Phi G^{[2]}(\lambda_{0}^{[k]})y^{[k]}=y^{[k]}$ is not appropriate.

 On the contrary, assume that the dynamic system ${\rm\bf \Sigma}$ is not observable. Then, according to Theorem 1 and Lemma 3, as well as Lemma 2, there exist at least one complex number, denote it by $\lambda_{0}$, and one nonzero complex vector, denote by $y_{0}$, such that
\begin{equation}
G(\lambda_{0})y_{0}=0
\label{eqn:a13}
\end{equation}

From Equation (\ref{eqn:a12}) and the definition of the TFM $G^{[1]}(\lambda)$, this means that ${\rm\bf diag}\{G^{[1]}_{i}(\lambda_{0})|_{i=1}^{N}\}y_{0}=0$.
Partition the vector $y_{0}$ as $y_{0}={\rm\bf col}\{y_{0i}|_{i=1}^{N}\}$ with the column vector $y_{0i}$ having a dimension of $m_{{\rm\bf S}i}$. Then, there exists at least one integer $i$ simultaneously satisfying $1\leq i\leq N$ and $y_{0i}\neq 0$, denote it by $i_{0}$, such that
\begin{equation}
G^{[1]}_{i_{0}}(\lambda_{0})y_{0i_{0}}=0
\end{equation}
Therefore, $\lambda_{0}$ is also a transmission zero of the TFM $G^{[1]}_{i_{0}}(\lambda)$ and $y_{0i_{0}}\in {\cal Y}_{i_{0}}^{[*]}$, in which $*$ is an integer belonging to the set $\{1,\;2,\;\cdots,\;m\}$. Hence,
\begin{equation}
y_{0}\in {\cal Y}^{[*]},\hspace{0.5cm}*\in\{1,2,\cdots,m\}
\end{equation}

On the other hand, from Equations (\ref{eqn:a12}) and (\ref{eqn:a13}), we have that $\left[\Phi G^{[2]}(\lambda_{0})-I_{M_{\rm\bf S}}\right]y_{0}=0$.
Hence
\begin{equation}
\Phi G^{[2]}(\lambda_{0}^{[*]})y_{0}=y_{0}
\end{equation}

That is, non-observability of the dynamic system ${\rm\bf\Sigma}$ certainly leads to the existence of an integer $k$ and a nonzero complex vector $y$ satisfy simultaneously $1\leq k\leq m$, $y\in{\cal Y}^{[k]}$ and $\Phi G^{[2]}(\lambda_{0}^{[k]})y=y$.

This completes the proof.  \hspace{\fill}$\Diamond$

\hspace*{-0.4cm}{\bf Proof of Corollary 1:} From the definitions of the matrices $A$ and $B$, which are given immediately after Equation (\ref{eqn:10}), it is clear that both of them only have real elements. It can therefore be declared that $x^{H}B\neq 0$ for every nonzero $x$ satisfying $x^{H}A=\lambda x^{H}$ is equivalent to that for every nonzero $x$ satisfying $A^{T}x=\lambda x$, $B^{T}x\neq 0$. On the basis of Lemma 1, it can be further declared that controllability of the matrix pair $(A,\;B)$ is equivalent to the observability of the matrix pair $(A^{T},\;B^{T})$.

On the other hand, from the definitions of the matrices $A$ and $B$, we have
\begin{eqnarray}
A^{T}&=&\left\{A_{\rm\bf
TT}\!+\!A_{\rm\bf TS}\Phi\left[I\!-\!A_{\rm\bf
SS}\Phi\right]^{-1}A_{\rm\bf ST}\right\}^{T} \nonumber\\
&=& A_{\rm\bf
TT}^{T}\!+\!A^{T}_{\rm\bf ST}\Phi^{T}\left[I\!-\!A^{T}_{\rm\bf
SS}\Phi^{T}\right]^{-1}A_{\rm\bf TS}^{T} \\
B^{T}&=&\left\{B_{\rm\bf
T}\!+\!A_{\rm\bf TS}\Phi\left[I\!-\!A_{\rm\bf
SS}\Phi\right]^{-1}B_{\rm\bf S}\right\}^{T}\nonumber\\
&=&B_{\rm\bf
T}^{T}\!+\!B_{\rm\bf S}^{T}\Phi^{T}\left[I\!-\!A_{\rm\bf
SS}^{T}\Phi^{T}\right]^{-1}A^{T}_{\rm\bf TS}
\end{eqnarray}
which have completely the same forms respectively as those of the matrix $A$ and the matrix $C$.

The proof can now be completed through directly utilizing Theorem 2.  \hspace{\fill}$\Diamond$

\hspace*{-0.4cm}{\bf Proof of Corollary 2:} Assume that for an arbitrary complex number $\lambda$ and an arbitrary nonzero complex vector $x$ satisfying simultaneously $|\lambda|\geq 1$ and $Ax=\lambda x$, $Cx\neq 0$. Assume also that there is an element in the set $\rm\bf\Lambda$, denote it by $\lambda_{0}^{[k]}$, satisfying $|\lambda_{0}^{[k]}|\geq 1$ such that there exists a
$y^{[k]}\in{\cal Y}^{[k]}$ with the property that
$\Phi {G}^{[2]}(\lambda_{0}^{[k]})y^{[k]}=y^{[k]}$. According to the definitions of the set ${\cal Y}^{[k]}$ and the TFM $G(\lambda)$, we have that
\begin{equation}
G(\lambda_{0}^{[k]})y^{[k]}=0
\label{eqn:36}
\end{equation}

Define a vector $w^{[k]}$ as $w^{[k]}=(\lambda_{0}^{[k]}I_{M_{\rm\bf T}}-A_{\rm\bf TT})^{-1}A_{\rm\bf TS}y^{[k]}$. As the matrix $\lambda I_{M_{\rm\bf T}}-A_{\rm\bf TT}$ is of a full normal rank, it can be declared that the vector $w^{[k]}$ is well defined. Then, Equation (\ref{eqn:36}) can be reexpressed as
\begin{equation}
M(\lambda_{0}^{[k]}){\rm\bf col}\{w^{[k]},\;y^{[k]}\}=0
\label{eqn:37}
\end{equation}

As the invertibility of the matrix $I_{M_{\rm\bf S}}-\Phi A_{\rm\bf SS}$ is guaranteed by the well-posedness of the dynamic system $\rm\bf\Sigma$, the above equation leads to
\begin{equation}
y^{[k]}=(I_{M_{\rm\bf S}}-\Phi A_{\rm\bf SS})^{-1}\Phi A_{\rm\bf ST}w^{[k]}
\label{eqn:38}
\end{equation}
Recalling that $y^{[k]}$ is not a zero vector, it is clear from this equality that the vector $w^{[k]}$ is not equal to zero also.

Substitute Equation (\ref{eqn:38}) back into Equation (\ref{eqn:37}), direct algebraic manipulations show that
\begin{equation}
\lambda_{0}^{[k]}w^{[k]}=Aw^{[k]},\hspace{0.5cm} Cw^{[k]}=0
\end{equation}

This is a contradiction, and hence the existence of such a $\lambda_{0}^{[k]}$ is impossible.

On the contrary, assume that there exist a complex number $\lambda_{0}$ and a nonzero complex vector $y_{0}$ such that $|\lambda_{0}|\geq 1$, $Ay_{0}=\lambda_{0} y_{0}$ and $Cy_{0}=0$ are satisfied simultaneously. Define a vector $z_{0}$ as $z_{0}=(I_{M_{\rm\bf S}}-\Phi A_{\rm\bf SS})^{-1}\Phi A_{\rm\bf ST}y_{0}$. Then, we have that
\begin{equation}
M(\lambda_{0}){\rm\bf col}\{y_{0},\;z_{0}\}=0
\label{eqn:39}
\end{equation}
Therefore, if the matrix $\lambda_{0}I_{M_{\rm\bf T}}-A_{\rm\bf TT}$ is invertible, then, $y_{0}=(\lambda_{0}I_{M_{\rm\bf T}}-A_{\rm\bf TT})^{-1}A_{\rm\bf TS}z_{0}$, $z_{0}\neq 0$ and $G(\lambda_{0})z_{0}=0$. This means that $\lambda_{0}\in{\rm\bf\Lambda}$, and there is a $k\in\{1,2,\cdots,m\}$, such that $z_{0}\in{\cal Y}^{[k]}$ and $\Phi {G}^{[2]}(\lambda_{0})z_{0}=z_{0}$.

Assume that the matrix $\lambda_{0}I_{M_{\rm\bf T}}-A_{\rm\bf TT}$ is rank deficient. As Equation (\ref{eqn:39}) means that $\left(\lambda_{0}I_{M_{\rm\bf T}}-A_{\rm\bf TT}\right)y_{0}$ $=A_{\rm\bf TS}z_{0}$, it is clear that $A_{\rm\bf TS}z_{0}$ belongs to the space spanned by the columns of the matrix $\lambda_{0}I_{M_{\rm\bf T}}-A_{\rm\bf TT}$. Therefore, there always exists a solution $y_{0}$ to this equation. On the other hand, as the matrix $A_{\rm\bf TT}$ is of a finite dimension, its eigenvalues must take some isolated values \cite{hj91}. This means that there certainly exists a positive number $\varepsilon$ such that for all $\delta\in\{-\varepsilon,\;\varepsilon\}/\{0\}$, $(\lambda_{0}-\delta)I_{M_{\rm\bf T}}-A_{\rm\bf TT}$ is invertible. Define the inverse of the matrix $\lambda_{0}I_{M_{\rm\bf T}}-A_{\rm\bf TT}$ as
\begin{equation}
\left(\lambda_{0}I_{M_{\rm\bf T}}-A_{\rm\bf TT}\right)^{-1}=\lim_{\delta\rightarrow 0^{+}}\left[(\lambda_{0}-\delta)I_{M_{\rm\bf T}}-A_{\rm\bf TT}\right]^{-1}
\end{equation}
Then, it can be proved that
\begin{equation}
y_{0}=(\lambda_{0}I_{M_{\rm\bf T}}-A_{\rm\bf TT})^{-1}A_{\rm\bf TS}z_{0}
\end{equation}
which further leads to the existence of an integer $k$ belonging to $\{1,2,\cdots,m\}$ and simultaneously satisfying $\lambda_{0}^{[k]}=\lambda_{0}$, $z_{0}\in{\cal Y}^{[k]}$ and $\Phi {G}^{[2]}(\lambda_{0})z_{0}=z_{0}$

The above arguments mean that the existence of a complex number $\lambda$ and a nonzero complex vector $y$ simultaneously satisfying  $|\lambda|\geq 1$, $Ay=\lambda y$ and $Cy=0$ is equivalent to the existence of an element $\lambda_{0}^{[k]}$ satisfying simultaneously $\lambda_{0}^{[k]}\in{\rm\bf\Lambda}$ and $|\lambda_{0}^{[k]}|\geq 1$ and guaranteing the availability of a $y^{[k]}\in{\cal Y}^{[k]}$ with the property that
$\Phi {G}^{[2]}(\lambda_{0}^{[k]})y^{[k]}=y^{[k]}$.

Similarly, it can also be proved that the existence of a complex number $\lambda$ and a nonzero complex vector $x$ simultaneously satisfying  $|\lambda|=1$, $x^{H}A=\lambda x^{H}$ and $x^{H}B=0$ is equivalent to the existence of an integer $\bar{k}$ satisfying simultaneously $\bar{\lambda}_{0}^{[k]}\in\bar{\rm\bf\Lambda}$  and $|\bar{\lambda}_{0}^{[\bar{k}]}|=1$ and guaranteing the availability of a $\bar{y}^{[\bar{k}]}\in\bar{\cal Y}^{[\bar{k}]}$ with the property $\Phi^{T} \bar{G}^{[2]}(\bar{\lambda}_{0}^{[\bar{k}]})\bar{y}^{[\bar{k}]}= \bar{y}^{[\bar{k}]}$.

The proof can now be completed utilizing Lemma 4.  \hspace{\fill}$\Diamond$

\hspace*{-0.4cm}{\bf Proof of Lemma 6:} According to the theory of Kalman filtering \cite{ksh}, if the covariance matrix $P^{[kal]}(t)$ is invertible, then,
\begin{equation}
P^{[kal]}(t+1)=A(t)\left[(P^{[kal]}(t))^{-1}+C^{T}(t)\left[D(t)D^{T}(t)\right]^{-1}C(t)\right]^{-1}A^{T}(t)+
B_{\rm\bf T}(t)B_{\rm\bf T}^{T}(t)
\label{eqn:a16}
\end{equation}

From Lemma 5 and the above the recursive formula for the covariance matrix of the Kalman filter, we further have that
\begin{eqnarray}
P_{ii}(t+1)-P_{ii}^{[kal]}(t+1)&=& A_{i}(t)\left[P^{-1}(t)+\bar{C}_{i}^{T}(t)\bar{C}_{i}(t)\right]^{-1}\!\!A_{i}^{T}(t)+B_{\rm\bf
T}(t,i)B_{\rm\bf T}^{T}(t,i)-\nonumber\\
& & \hspace*{0.8cm}J_{{\rm\bf
T}i}^{T}\left\{
A(t)\left[(P^{[kal]})^{-1}(t)+\bar{C}^{T}(t)\bar{C}(t)\right]^{-1}\!\!A^{T}(t)+B_{\rm\bf
T}(t)B_{\rm\bf T}^{T}(t)\right\}J_{{\rm\bf
T}i}\nonumber\\
&=& A_{i}(t)\left\{\left[P^{-1}(t)+\bar{C}_{i}^{T}(t)\bar{C}_{i}(t)\right]^{-1}-
\left[(P^{[kal]})^{-1}(t)+\bar{C}^{T}(t)\bar{C}(t)\right]^{-1}\right\}
\!\!A_{i}^{T}(t)\nonumber\\
&=& A_{i}(t)\left[(P^{[kal]})^{-1}(t)+\bar{C}^{T}(t)\bar{C}(t)\right]^{-1}\left\{\left[(P^{[kal]})^{-1}(t)+\bar{C}^{T}(t)\bar{C}(t)\right]-\right.\nonumber\\
& & \hspace*{0.8cm} \left.\left[P^{-1}(t)+\bar{C}_{i}^{T}(t)\bar{C}_{i}(t)\right]\right\}
\left[P^{-1}(t)+\bar{C}_{i}^{T}(t)\bar{C}_{i}(t)\right]^{-1}
\!\!A_{i}^{T}(t)\nonumber\\
&=& A_{i}(t)\left[(P^{[kal]})^{-1}(t)+\sum_{k=1}^{N}\bar{C}_{j}^{T}(t)\bar{C}_{j}(t)\right]^{-1}\left\{(P^{[kal]})^{-1}(t)-P^{-1}(t)+\right.\nonumber\\
& & \hspace*{0.8cm}\left.
\sum_{k=1,k\neq i}^{N}\bar{C}_{j}^{T}(t)\bar{C}_{j}(t)\right\}
\left[P^{-1}(t)+\bar{C}_{i}^{T}(t)\bar{C}_{i}(t)\right]^{-1}
\!\!A_{i}^{T}(t)
\end{eqnarray}

This completes the proof.  \hspace{\fill}$\Diamond$

\hspace*{-0.4cm}{\bf Proof of Theorem 3:} When $P(t)=P^{[kal]}(t)$, results of Lemma 6 directly lead to
\begin{eqnarray}
& & P_{ii}(t+1)-P_{ii}^{[kal]}(t+1) \nonumber\\
&=&
A_{i}(t)\left[P^{-1}(t)+\bar{C}^{T}(t)\bar{C}(t)\right]^{-1}\left\{\bar{C}^{T}(t)\bar{C}(t)-
\bar{C}_{i}^{T}(t)\bar{C}_{i}(t)\right\}
\left[P^{-1}(t)+\bar{C}_{i}^{T}(t)\bar{C}_{i}(t)\right]^{-1}
\!\!A_{i}^{T}(t)\nonumber\\
&=& A_{i}(t)\left\{\left[P^{-1}(t)+\bar{C}_{i}^{T}(t)\bar{C}_{i}(t)\right]^{-1}-\left[P^{-1}(t)+\bar{C}^{T}(t)\bar{C}(t)\right]^{-1}\right\}A_{i}^{T}(t)
\label{eqn:b1}
\end{eqnarray}

For brevity, define matrices $X_{i}(t)$ and $\hat{C}_{i}(t)$ respectively as $X_{i}(t)=\left[P^{-1}(t)+\bar{C}_{i}^{T}(t)\bar{C}_{i}(t)\right]^{1/2}$ and
$\hat{C}_{i}(t)={\rm\bf col}\{\bar{C}_{j}(t)|_{j=1,j\neq i}^{N}\}$. Then,
\begin{eqnarray}
& & \left[P^{-1}(t)+\bar{C}_{i}^{T}(t)\bar{C}_{i}(t)\right]^{-1}-\left[P^{-1}(t)+\bar{C}^{T}(t)\bar{C}(t)\right]^{-1}\nonumber\\
&=& X_{i}^{-2}(t)-\left[X_{i}^{2}(t)+
\hat{C}_{i}^{T}(t)\hat{C}_{i}(t)\right]^{-1} \nonumber\\
&=& X_{i}^{-1}(t)\left\{I-\left[I+X_{i}^{-1}(t)\hat{C}_{i}^{T}(t)\hat{C}_{i}(t)X_{i}^{-1}(t)\right]^{-1}\right\}X_{i}^{-1}(t)\nonumber\\
&=& X_{i}^{-1}(t)\left[I+X_{i}^{-1}(t)\hat{C}_{i}^{T}(t)\hat{C}_{i}(t)X_{i}^{-1}(t)\right]^{-1}X_{i}^{-1}(t)\hat{C}_{i}^{T}(t)\hat{C}_{i}(t)X_{i}^{-2}(t)\nonumber\\
&=& X_{i}^{-2}(t)\hat{C}_{i}^{T}(t)\left[I+\hat{C}_{i}(t)X_{i}^{-2}(t)\hat{C}_{i}^{T}(t)\right]^{-1}\hat{C}_{i}(t)X_{i}^{-2}(t)
\label{eqn:b2}
\end{eqnarray}

Therefore,
\begin{eqnarray}
P_{ii}(t+1)-P_{ii}^{[kal]}(t+1)&=& A_{i}(t)X_{i}^{-2}(t)\hat{C}_{i}^{T}(t)\left[I+\hat{C}_{i}(t)X_{i}^{-2}(t)\hat{C}_{i}^{T}(t)\right]^{-1}\hat{C}_{i}(t)X_{i}^{-2}(t)A_{i}^{T}(t)
\nonumber\\
&=& \left\{A_{i}(t)X_{i}^{-2}(t)\hat{C}_{i}^{T}(t)\left[I+\hat{C}_{i}(t)X_{i}^{-2}(t)\hat{C}_{i}^{T}(t)\right]^{-1/2}\right\}\left\{\star\right\}^{T}
\label{eqn:b3}
\end{eqnarray}

Hence, $P_{ii}(t+1)-P_{ii}^{[kal]}(t+1)=0$ if and only if
\begin{equation}
A_{i}(t)X_{i}^{-2}(t)\hat{C}_{i}^{T}(t)\left[I+\hat{C}_{i}(t)X_{i}^{-2}(t)\hat{C}_{i}^{T}(t)\right]^{-1/2}=0
\end{equation}
which is further equivalent to
\begin{equation}
A_{i}(t)\left[P^{-1}(t)+\bar{C}_{i}^{T}(t)\bar{C}_{i}(t)\right]^{-1}\hat{C}_{i}^{T}(t)=0
\label{eqn:a14}
\end{equation}

Note that
\begin{eqnarray}
& & \left[P^{-1}(t)+\bar{C}_{i}^{T}(t)\bar{C}_{i}(t)\right]^{-1}\hat{C}_{i}^{T}(t) \nonumber\\
&=& \left[P^{-1}(t)+\bar{C}^{T}(t)\bar{C}(t)-\hat{C}_{i}^{T}(t)\hat{C}_{i}(t)\right]^{-1}\hat{C}_{i}^{T}(t)\nonumber\\
&=& \left[P^{-1}(t)+\bar{C}^{T}(t)\bar{C}(t)\right]^{-1}\left\{I-\hat{C}_{i}^{T}(t)\hat{C}_{i}(t)\left[P^{-1}(t)+\bar{C}^{T}(t)\bar{C}(t)\right]\right\}^{-1}\hat{C}_{i}^{T}(t)\nonumber\\
&=& \left[P^{-1}(t)+\bar{C}^{T}(t)\bar{C}(t)\right]^{-1}\hat{C}_{i}^{T}(t)\left\{I-\hat{C}_{i}(t)\left[P^{-1}(t)+\bar{C}^{T}(t)\bar{C}(t)\right]\hat{C}_{i}^{T}(t)\right\}^{-1}
\label{eqn:a15}
\end{eqnarray}

Combining Equations (\ref{eqn:a14}) and (\ref{eqn:a15}) together, it can be claimed that a necessary and sufficient condition for the satisfaction of $P_{ii}(t+1)-P_{ii}^{[kal]}(t+1)=0$ is that
\begin{equation}
A_{i}(t)\left[P^{-1}(t)+\bar{C}^{T}(t)\bar{C}(t)\right]^{-1}\hat{C}_{i}^{T}(t)=0
\label{eqn:a17}
\end{equation}

From the definition of the matrix $\hat{C}_{i}(t)$, it is straightforward to see that this condition is equivalent to
\begin{equation}
A_{i}(t)\left[P^{-1}(t)+\bar{C}^{T}(t)\bar{C}(t)\right]^{-1}\bar{C}_{j}^{T}(t)=0, \hspace{0.5cm} \forall j\neq i
\label{eqn:a18}
\end{equation}

Note that according to its definition, $\bar{C}_{j}(t)=J_{{\rm\bf y}\!j}^{T}\bar{C}(t)$. Direct matrix manipulations show that Equation (\ref{eqn:a18}) can be rewritten as
\begin{equation}
A_{i}(t)P(t)\bar{C}^{T}(t)[I+\bar{C}(t)P(t)\bar{C}^{T}(t)]^{-1}J_{{\rm\bf
y}\!j}=0
\end{equation}

On the other hand, from Lemma 5 and Equation (\ref{eqn:a16}),
we have that when $P(t)=P^{[kal]}(t)$ and $i\neq j$,
\begin{eqnarray}
& & P_{ij}(t+1)-P_{ij}^{[kal]}(t+1) \nonumber\\
&=& A_{i}(t)\left[P^{-1}(t)+\bar{C}_{i}^{T}(t)\bar{C}_{i}(t)\right]^{-1}P^{-1}(t)\left[P^{-1}(t)+\bar{C}_{j}^{T}(t)\bar{C}_{j}(t)\right]^{-1}\!\!A_{j}^{T}(t)-\nonumber\\
& & \hspace*{0.8cm}J_{{\rm\bf
T}i}^{T}\left\{
A(t)\left[P^{-1}(t)+\bar{C}^{T}(t)\bar{C}(t)\right]^{-1}\!\!A^{T}(t)+B_{\rm\bf
T}(t)B_{\rm\bf T}^{T}(t)\right\}J_{{\rm\bf
T}j}\nonumber\\
&=& A_{i}(t)\left[P^{-1}(t)+\bar{C}^{T}(t)\bar{C}(t)\right]^{-1}\left\{\left[P^{-1}(t)+\bar{C}_{i}^{T}(t)\bar{C}_{i}(t)+\hat{C}_{i}^{T}(t)\hat{C}_{i}(t)\right]\times\right.\nonumber\\
& & \hspace*{0.8cm}\left.\left[P^{-1}(t)+\bar{C}_{i}^{T}(t)\bar{C}_{i}(t)\right]^{-1}P^{-1}(t)\left[P^{-1}(t)+\bar{C}_{j}^{T}(t)\bar{C}_{j}(t)\right]^{-1}-I\right\}
\!\!A_{j}^{T}(t) \nonumber\\
&=& A_{i}(t)\left[P^{-1}(t)+\bar{C}^{T}(t)\bar{C}(t)\right]^{-1}\left\{\left[I+\hat{C}_{i}^{T}(t)\hat{C}_{i}(t)\left[P^{-1}(t)+\bar{C}_{i}^{T}(t)\bar{C}_{i}(t)\right]^{-1}\right]P^{-1}(t)\right.\nonumber\\
& & \hspace*{0.8cm}\left.-\left[P^{-1}(t)+\bar{C}_{j}^{T}(t)\bar{C}_{j}(t)\right]\right\}\left[P^{-1}(t)+\bar{C}_{j}^{T}(t)\bar{C}_{j}(t)\right]^{-1}
\!\!A_{j}^{T}(t) \nonumber\\
&=& A_{i}(t)\left[P^{-1}(t)+\bar{C}^{T}(t)\bar{C}(t)\right]^{-1}\hat{C}_{i}^{T}(t)\hat{C}_{i}(t)\left[P^{-1}(t)+\bar{C}_{i}^{T}(t)\bar{C}_{i}(t)\right]^{-1}P^{-1}(t)\times\nonumber\\
& & \hspace*{3cm}\left[P^{-1}(t)+\bar{C}_{j}^{T}(t)\bar{C}_{j}(t)\right]^{-1}
\!\!A_{j}^{T}(t)\nonumber\\
& & \hspace*{0.8cm}-A_{i}(t)\left[P^{-1}(t)+\bar{C}^{T}(t)\bar{C}(t)\right]^{-1}\bar{C}_{j}^{T}(t)\bar{C}_{j}(t)\left[P^{-1}(t)+\bar{C}_{j}^{T}(t)\bar{C}_{j}(t)\right]^{-1}
\!\!A_{j}^{T}(t)
\end{eqnarray}

From Equations (\ref{eqn:a17}) and (\ref{eqn:a18}), we have that if $P_{ii}(t+1)=P_{ii}^{[kal]}(t+1)$, then, both the matrix $A_{i}(t)\left[P^{-1}(t)+\bar{C}^{T}(t)\bar{C}(t)\right]^{-1}\hat{C}_{i}^{T}(t)$ and the matrix $A_{i}(t)\left[P^{-1}(t)+\bar{C}^{T}(t)\bar{C}(t)\right]^{-1}\bar{C}_{j}^{T}(t)$ are equal to zero. Therefore, when the $i$-th diagonal block of the covariance matrix of the CDOSSP is equal to that of the Kalman filter, it is certain that the other blocks in the same row are also equal to its counterpart. That is, for any $j$ belonging to the set $\{1,\;2,\;\cdots,\;N\}/\{i\}$,
\begin{equation}
P_{ij}(t+1)=P_{ij}^{[kal]}(t+1)
\end{equation}

In addition, from the theory of Kalman filtering \cite{ksh,zhou10b,zhou11}, we have that its update gain matrix can be expressed as
\begin{equation}
K^{[kal]}(t)=A(t)\left[(P^{[kal]}(t))^{-1}+\bar{C}^{T}(t)\bar{C}(t)\right]^{-1}\bar{C}^{T}(t)
\end{equation}
Then, from the definition of the matrices $J_{{\rm\bf
T}i}$ and $J_{{\rm\bf
y}i}$, we have that its $i$-th row $j$-th column block, that is,  $K_{ij}^{[kal]}(t)$, can be further rewritten as
\begin{eqnarray}
K_{ij}^{[kal]}(t)&=& J_{{\rm\bf
T}i}^{T} \left\{A(t)\left[(P^{[kal]}(t))^{-1}+\bar{C}^{T}(t)\bar{C}(t)\right]^{-1}\bar{C}^{T}(t)\right\}J_{{\rm\bf
y}\!j} \nonumber\\
&=& A_{i}(t)\left[(P^{[kal]}(t))^{-1}+\bar{C}^{T}(t)\bar{C}(t)\right]^{-1}\bar{C}_{j}^{T}(t)
\end{eqnarray}

It can therefore be declared from Equation (\ref{eqn:a18}) that if $P_{ii}(t+1)=P_{ii}^{[kal]}(t+1)$, then
\begin{equation}
K_{ij}^{[kal]}(t)=\left\{\begin{array}{ll}
A_{i}(t)\left[(P^{[kal]}(t))^{-1}+\bar{C}^{T}(t)\bar{C}(t)\right]^{-1}\bar{C}_{i}^{T}(t) & i=j \\
0 & i\neq j
\end{array}\right.
\end{equation}

This completes the proof.  \hspace{\fill}$\Diamond$

\hspace*{-0.4cm}{\bf Proof of Theorem 4:} Assume that the CDOSSP and the Kalman filter have the same steady state covariance matrix of estimation errors. That is,
\begin{equation}
\lim_{t\rightarrow\infty}P(t)=\lim_{t\rightarrow\infty}P^{[kal]}(t)=P
\end{equation}
Then, for arbitrary $i=1,2,\cdots,N$,
\begin{equation}
\lim_{t\rightarrow\infty}\left[P_{ii}(t)-P_{ii}^{[kal]}(t)\right]=0
\end{equation}

On the other hand, according to Lemma 6 and
Equations (\ref{eqn:b1})-(\ref{eqn:b3}), we have that
\begin{eqnarray}
& &
\lim_{t\rightarrow\infty}\left[P_{ii}(t+1)-{P}_{ii}^{[kal]}(t+1)\right]\nonumber\\
&=&\lim_{t\rightarrow\infty}A_{i}\left\{[{P}^{[kal]}(t)]^{-1}\!+\!\bar{C}^{T}\bar{C}\right\}^{\!-1}
\!\!\left\{[{P}^{[kal]}(t)]^{\!-1}\!-\!P^{-1}(t)\!+\!\bar{C}^{T}\bar{C}\!-\!\bar{C}_{i}^{T}\bar{C}_{i}\right\}\![P^{-1}(t)\!+\!\bar{C}_{i}^{T}\bar{C}_{i}]^{\!-1}A_{i}^{T}\nonumber\\
&=& A_{i}\left\{P^{-1}+\bar{C}^{T}\bar{C}\right\}^{-1}
\left\{P^{-1}-P^{-1}+\bar{C}^{T}\bar{C}-\bar{C}_{i}^{T}\bar{C}_{i}\right\}[P^{-1}+\bar{C}_{i}^{T}\bar{C}_{i}]^{-1}A_{i}^{T}\nonumber\\
&=&\left\{A_{i}[P^{-1}+\bar{C}_{i}^{T}\bar{C}_{i}]^{-1}\hat{C}_{i}^{T}\left[I+\hat{C}_{i}(P^{-1}+\bar{C}_{i}^{T}\bar{C}_{i})^{-1}\hat{C}_{i}^{T}\right]^{-1/2}\right\}\left\{\star\right\}^{T}
\end{eqnarray}

Therefore, these two state estimators have the same steady state estimation accuracy only if
\begin{equation}
A_{i}[P^{-1}+\bar{C}_{i}^{T}\bar{C}_{i}]^{-1}\hat{C}_{i}^{T}=0,\hspace{0.5cm} i=1,2,\cdots, N
\label{eqn:b4}
\end{equation}

Based on the same arguments as those in Equations (\ref{eqn:a15})-(\ref{eqn:a18}), it can be proved that this equation is equivalent to
\begin{equation}
A_{i}P\bar{C}^{T}[I+\bar{C}P\bar{C}^{T}]^{-1}J_{{\rm\bf y}\!j}=0,\hspace{0.5cm} i,j=1,2,\cdots, N; \;\;i\neq j
\label{eqn:b5}
\end{equation}

On the contrary, assume that the dynamic system $\bar{\rm\bf\Sigma}$ is time invariant and its system matrices satisfy Equation (\ref{eqn:b5}). Then, using completely the same arguments as those in the proof of Theorem 3, it can be proved that for arbitrary $i,j\in\{1,2,\cdots,N\}$ with $j\neq i$,
\begin{equation}
\lim_{t\rightarrow\infty}K^{[kal]}_{ij}(t)=0
\label{eqn:b6}
\end{equation}
That is, in its steady state, the update gain matrix of the Kalman filter is block diagonal.

Initialize both the Kalman filter and the CDOSSP with the steady state covariance matrix of the Kalman filter. Then, according to Theorem 3 and Equation (\ref{eqn:b6}), satisfaction of Equation (\ref{eqn:40}) means that the update gain matrix of the Kalman filter is always block diagonal.

Note that the update gain matrix of the CDOSSP is proved to be the optimal one among all the block diagonal update gain matrices. It can therefore be declared that the covariance matrix of its estimation errors must not be greater than that of the Kalman filter. On the other hand, Kalman filter is proved to be the optimal estimator for linear plants with normal external disturbances \cite{ksh,simon06,zhou10b,zhou11}, its estimation accuracy must not be lower than the CDOSSP when measured through covariances. These mean that these two state estimators must have the same covariance matrix at all the time instants. As the Kalman filter is assumed to be convergent, the CDOSSP must also converge with this initial condition. Therefore, these two predictors must have the same covariance matrix of prediction errors at their steady states.

The above arguments can be easily modified to situations in which the Kalman filter takes another initial covariance matrix, which essentially only requires an appropriate transformation between the temporal variables of these two predictors. As a matter of fact, let $t_{k}$ and $t_{c}$ denote the temporal variables respectively of the Kalman filter and the CDOSSP, and assign $t_{c}$ as $t_{c}=t_{k}+\delta_{t}$. Then, on the basis of Theorem 3 and taking the limit of $\delta_{t}\rightarrow\infty$, completely the same results can be established through similar arguments. The details are omitted due to their obviousness.

This completes the proof.  \hspace{\fill}$\Diamond$

\end{document}